\documentclass[english]{article}
\usepackage[T1]{fontenc}
\usepackage[utf8]{inputenc}
\usepackage{geometry}
\usepackage{babel}
\usepackage{mathtools}
\usepackage{amsmath}
\usepackage{amsthm}
\usepackage{amssymb}
\usepackage{graphicx}
\usepackage[unicode=true,pdfusetitle,
 bookmarks=true,bookmarksnumbered=false,bookmarksopen=false,
 breaklinks=false,pdfborder={0 0 1},backref=false,colorlinks=false]
 {hyperref}

\makeatletter
\theoremstyle{definition}
\newtheorem{defn}{\protect\definitionname}[section]
\theoremstyle{plain}
\newtheorem*{thm*}{\protect\theoremname}
\theoremstyle{plain}
\newtheorem*{lem*}{\protect\lemmaname}
\theoremstyle{plain}
\newtheorem{conjecture}{\protect\conjecturename}[section]
\theoremstyle{remark}
\newtheorem*{rem*}{\protect\remarkname}
\theoremstyle{plain}
\newtheorem{lem}{\protect\lemmaname}[section]
\theoremstyle{plain}
\newtheorem{cor}{\protect\corollaryname}[section]
\theoremstyle{plain}
\newtheorem{thm}{\protect\theoremname}[section]
\theoremstyle{remark}
\newtheorem*{claim*}{\protect\claimname}
\theoremstyle{definition}
\newtheorem{example}{\protect\examplename}[section]
\theoremstyle{plain}
\newtheorem*{cor*}{\protect\corollaryname}

\usepackage[algo2e,linesnumbered,ruled]{algorithm2e}

\usepackage{tikz}
\usetikzlibrary{calc}

\usepackage{mathbbol}
\def\bbone{\mathbb{1}}

\newcommand{\Bern}{\mathrm{Bern}}
\newcommand{\Binom}{\mathrm{Binom}}
\newcommand{\Exp}{\mathrm{Exp}}
\newcommand{\Pois}{\mathrm{Pois}}

\newcommand{\conv}{\mathrm{conv}}

\newcommand{\CG}{\mathrm{CG}}

\DeclareMathOperator{\supp}{supp}

\@ifundefined{showcaptionsetup}{}{%
 \PassOptionsToPackage{caption=false}{subfig}}
\usepackage{subfig}
\makeatother

\providecommand{\claimname}{Claim}
\providecommand{\conjecturename}{Conjecture}
\providecommand{\corollaryname}{Corollary}
\providecommand{\definitionname}{Definition}
\providecommand{\examplename}{Example}
\providecommand{\lemmaname}{Lemma}
\providecommand{\remarkname}{Remark}
\providecommand{\theoremname}{Theorem}

\begin{document}
\begin{center}
\includegraphics[height=1.5cm]{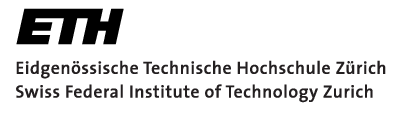}\vspace{5cm}
\par\end{center}

\begin{center}
{\large Ivan Sergeev}\vspace{1.5cm}
\par\end{center}

\begin{center}
{\Huge \bf On contention resolution for the hypergraph matching, knapsack, and $k$-column sparse packing problems \par}\vspace{2.5cm}
{\large \bf Master's Thesis}\\[3ex]
Swiss Federal Institute of Technology (ETH) Zurich\\\vspace{3.5cm}
{\bf Supervision} \\[1.5ex]
Prof. Dr. Rico Zenklusen \\
Dr. Charalampos Angelidakis\vfill{}
March 2, 2020
\thispagestyle{empty}
\par\end{center}

\newpage{}

\section*{Abstract}

The contention resolution framework is a versatile rounding technique
used as a part of the relaxation and rounding approach for solving
constrained submodular function maximization problems. We apply this
framework to the hypergraph matching, knapsack, and $k$-column sparse
packing problems. In the hypergraph matching setting, we adapt the
technique from \cite{Guruganesh2018} to non-constructively prove
that the correlation gap is at least $\frac{1-e^{-k}}{k}$ and provide
a monotone $\left(b,\frac{1-e^{-bk}}{bk}\right)$-balanced contention
resolution scheme, generalizing the results of \cite{Bruggmann2019}.
For the knapsack problem, we prove that the correlation gap of instances
where exactly $k$ copies of each item fit into the knapsack is at
least $\frac{1-e^{-2}}{2}$ and provide several monotone contention
resolution schemes: a $\frac{1-e^{-2}}{2}$-balanced scheme for instances
where all item sizes are strictly bigger than $\frac{1}{2}$, a $\frac{4}{9}$-balanced
scheme for instances where all item sizes are at most $\frac{1}{2}$,
and a $0.279$-balanced scheme for instances with arbitrary item sizes.
For $k$-column sparse packing integer programs, we slightly modify
the $\left(2k+o\left(k\right)\right)$-approximation algorithm for
$k$-CS-PIP based on the strengthened LP relaxation presented in \cite{Brubach2019}
to obtain a $\frac{1}{4k+o\left(k\right)}$-balanced contention resolution
scheme and hence a $\left(4k+o\left(k\right)\right)$-approximation
algorithm for $k$-CS-PIP based on the natural LP relaxation.

\section*{Acknowledgements}

I would like to thank my supervisors Rico Zenklusen and Charalampos
Angelidakis for introducing me to the topic and helpful discussions.
I am also grateful to Charalampos Angelidakis, Fedor Sergeev, and
Elias Wirth for detailed feedback, which helped improve the presentation
of the thesis.

\newpage{}

\tableofcontents{}

\newpage{}

\section{\label{sec:intro}Introduction}

\subsection{\label{subsec:intro cg}Contention resolution schemes and correlation
gap}

Contention resolution schemes, first introduced in \cite{Chekuri2014},
arise as a part of a general relaxation and rounding framework for
solving constrained submodular function maximization problems.
\begin{defn}[CSFM]
Let $E$ be a finite ground set, let $\mathcal{F}\subseteq2^{E}$
be a family of feasible sets, and let $f\colon2^{E}\to\mathbb{R}_{\ge0}$
be a submodular function. Then the maximization problem 
\[
\max_{S\in\mathcal{F}}f\left(S\right)
\]
is called the constrained submodular function maximization problem
(CSFM). 
\end{defn}
Submodular functions are very commonly used as utility functions to
describe diminishing returns, which is why CSFM is prominently applied,
for example, in algorithmic game theory. Some well-known problems
(e.g.,\,Max-Cut, Max-$k$-Cover) are examples of submodular function
maximization. The three most commonly used classes of methods for
solving CSFM are greedy algorithms, local search, and relaxation and
rounding approaches (see references listed in Section 1 of \cite{Bruggmann2019}).
As explained in Section 1 of \cite{Chekuri2014}, the former two combinatorial
techniques show limited success: they were unable to achieve optimal
approximation results except for some special cases (namely, a single
cardinality or knapsack constraint) and they do not allow to easily
combine constraints of different types. In constrast, the contention
resolution framework, which is a subclass of the relaxation and rounding
approaches, is more versatile, since it is closely related to polyhedral
techniques, can be applied to different types of constraints, and
allows to subdivide a problem into clean subproblems.

Before we introduce the general relaxation and rounding approach for
CSFM, we need to set up some additional notation. Throughout this
work, we assume $\mathcal{F}$ to be down-closed (as noted in Section
1 of \cite{Bruggmann2019}, this is without loss of generality if
$f$ is monotone). We denote the combinatorial polytope corresponding
to $\mathcal{F}$ as
\[
P_{\mathcal{F}}\coloneqq\conv\left(\left\{ \chi^{S}\mid S\in\mathcal{F}\right\} \right)\subseteq\left[0,1\right]^{E}.
\]
A polytope $P\subseteq\left[0,1\right]^{E}$ is called a relaxation
of $\mathcal{F}$ (or, equivalently, of $P_{\mathcal{F}}$) if $P\cap\left\{ 0,1\right\} ^{E}=P_{\mathcal{F}}\cap\left\{ 0,1\right\} ^{E}$,
i.e.,\,$P$ has the same integer points as $P_{\mathcal{F}}$. The
generic relaxation and rounding approach for CSFM uses a relaxation
$P$ of $\mathcal{F}$ and an extension $F\colon\left[0,1\right]^{E}\to\mathbb{R}_{\ge0}$
of $f$ and consists of two steps:
\begin{enumerate}
\item Find $x\in P$ maximizing $F$ over $P$.
\item Round $x$ to a point $\chi^{S}\in P$ and return $S$.
\end{enumerate}
The currently most successful extension $F$ is the multilinear extension
introduced in \cite{Calinescu2011}, which is defined as follows:
\[
F_{\mathrm{ML}}\left(x\right)\coloneqq\mathbb{E}\left[f\left(R\left(x\right)\right)\right]=\sum_{S\subseteq E}f\left(S\right)\prod_{e\in S}x_{e}\prod_{e\notin S}\left(1-x_{e}\right)
\]
for every $x\in\left[0,1\right]^{E}$, where $f\colon2^{E}\to\mathbb{R}_{\ge0}$
is the submodular function we want to extend, and $R\left(x\right)\subseteq E$
is a random subset containing each element $e\in E$ with probability
$x_{e}$. The success of this extension comes from its rich properties
that can be exploited in relaxation and rounding. More specifically,
there exist constant-factor approximation algorithms for maximizing
$F=F_{\mathrm{ML}}$ over any down-closed solvable polytope $P$,
i.e.,\,for the first step of the generic relaxation and rounding
approach (see references in Section 1 of \cite{Bruggmann2019}), whereas
for the second step, the most versatile rounding technique known currently
is the contention resolution scheme framework introduced in \cite{Chekuri2014},
which is closely related to the multilinear extension.
\begin{defn}[CRS, \cite{Chekuri2014}]
\label{def:intro crs}A $\left(b,c\right)$-balanced contention resolution
scheme (or CRS for short) $\pi$ for a relaxation $P$ of $\mathcal{F}$
is an algorithm that for every $x\in bP\coloneqq\left\{ by\mid y\in P\right\} $
and $A\subseteq\supp\left(x\right)$ returns a (possibly random) set
$\pi_{x}\left(A\right)\subseteq A$ such that $\pi_{x}\left(A\right)\in\mathcal{F}$
and
\[
\Pr\left[e\in\pi_{x}\left(R\left(x\right)\right)\mid e\in R\left(x\right)\right]=\frac{1}{x_{e}}\cdot\Pr\left[e\in\pi_{x}\left(R\left(x\right)\right)\right]\ge c\quad\forall e\in\supp\left(x\right),
\]
where $R\left(x\right)\subseteq E$ is a random set containing each
element $e\in E$ independently with probability $x_{e}$. The CRS
$\pi$ is monotone if $\Pr\left[e\in\pi_{x}\left(A\right)\right]\ge\Pr\left[e\in\pi_{x}\left(B\right)\right]$
for any $e\in A\subseteq B\subseteq\supp\left(x\right)$. For the
sake of simplicity, we call a CRS $c$-balanced if it is $\left(1,c\right)$-balanced.
\end{defn}
The terminology is inspired by fair contention resolution --- an
approach used to resolve conflicts between players' choices of items
in the allocation setting \cite{Feige2006,Feige2010}.

The utility of contention resolution schemes is summarized by the
following theorem.
\begin{thm*}[\cite{Chekuri2014}, Theorem 1.3]
Let $\pi$ be a monotone $c$-balanced CRS for a relaxation $P\subseteq\left[0,1\right]^{E}$
of $\mathcal{F}\subseteq2^{E}$ and let $x\in P$. Then
\[
\mathbb{E}\left[f\left(\pi_{x}\left(R\left(x\right)\right)\right)\right]\ge c\cdot F_{\mathrm{ML}}\left(x\right),
\]
where $F_{\text{ML}}$ is the multilinear extension of a monotone
submodular function $f\colon2^{E}\to\mathbb{R}_{\ge0}$. Moreover,
if $f$ is non-monotone, there exists an efficient procedure that
takes $\pi_{x}\left(R\left(x\right)\right)$ and returns a set $I_{x}\subseteq\pi_{x}\left(R\left(x\right)\right)$
with
\[
\mathbb{E}\left[f\left(I_{x}\right)\right]\ge c\cdot F_{\mathrm{ML}}\left(x\right).
\]
\end{thm*}
Combining contention resolution schemes for different constraints
is possible due to the following lemma.
\begin{lem*}[\cite{Chekuri2014}, Lemma 1.6]
Let $\mathcal{I}_{i}$, $i\in\left[\ell\right]$, be $\ell$ independence
systems and let $P_{\mathcal{I}_{i}}$ be the corresponding combinatorial
polytopes. Suppose there exists a monotone $\left(b,c_{i}\right)$-balanced
CRS for each polytope $P_{\mathcal{I}_{i}}$, $i\in\left[\ell\right]$.
Then the combinatorial polytope $P_{\mathcal{I}}$ for the independence
system $\mathcal{I}\coloneqq\bigcap_{i\in\left[\ell\right]}\mathcal{I}_{i}$
has a monotone $\left(b,\prod_{i\in\left[\ell\right]}c_{i}\right)$-balanced
CRS. Moreover, if every element of the ground set participates in
at most $k$ constraints and $c_{i}=c$ for every $i\in\left[\ell\right]$,
then there exists a monotone $\left(b,c^{k}\right)$-balanced CRS
for $P_{\mathcal{I}}$. In addition, if the CRS for every $P_{\mathcal{I}_{i}}$
is efficiently implementable, then so is the combined CRS for $P_{\mathcal{I}}$.
\end{lem*}
In this work, we also use the term contention resolution scheme when
referring to an algorithm that for every $x\in bP$ and $A\subseteq\supp\left(x\right)$
returns a point $y_{x}^{A}\in P_{\mathcal{F}}$ such that $\supp\left(y_{x}^{A}\right)\subseteq A$
and $\mathbb{E}\left[\left(y_{x}^{R\left(x\right)}\right)_{e}\right]\ge c\cdot x_{e}$
for every $e\in\supp\left(x\right)$. This is due to Proposition 11
from \cite{Bruggmann2019}, which states that any such algorithm can
be efficiently transformed into a $\left(b,c\right)$-balanced CRS
for $P$ and that if such an algorithm satisfies $\left(y_{x}^{A}\right)_{e}\ge\left(y_{x}^{B}\right)_{e}$
for every $e\in A\subseteq B\subseteq\supp\left(x\right)$, then the
resulting CRS is monotone.

A notion that is closely related to contention resolution schemes
is the correlation gap. In \cite{Bruggmann2019,Chekuri2014}, it is
defined it as follows.
\begin{defn}[Correlation gap]
\label{def:intro cg orig}For $\mathcal{F}\subseteq2^{E}$, the correlation
gap of $\mathcal{F}$ is defined as
\[
\CG\left(\mathcal{F}\right)\coloneqq\inf\left\{ \frac{\mathbb{E}\left[\max\left\{ \sum\limits _{e\in S}v_{e}\mid S\subseteq R\left(x\right),\ S\in\mathcal{F}\right\} \right]}{\sum\limits _{e\in E}v_{e}x_{e}}\mid x\in P_{\mathcal{F}},\ v\ge0\right\} ,
\]
where $R\left(x\right)\subseteq E$ is a random set containing each
element $e\in E$ independently with probability $x_{e}$. For a relaxation
$P$ of $\mathcal{F}$, the correlation gap of $P$ is defined by
replacing $P_{\mathcal{F}}$ with $P$ in the definition of $\CG\left(\mathcal{F}\right)$.
\end{defn}
The relationship between contention resolution schemes and the correlation
gap is summarized by the following theorem.
\begin{thm*}[\cite{Chekuri2014}, Theorem 4.6]
The correlation gap of $\mathcal{F}$ is equal to the maximum $c$
such that $\mathcal{F}$ admits a $c$-balanced CRS (but not necessarily
monotone).
\end{thm*}

\subsection{\label{sec:intro hg}Hypergraph matching problem}

This section focuses on the following optimization problem: find a
maximum weight matching in a weighted hypergraph, i.e.,\,a subset
of pairwise-disjoint hyperedges. We consider the case where each hyperedge
is incident to at most $k$ vertices. Given a hypergraph $H=\left(V,E\right)$
and a hyperedge weight vector $w\in\mathbb{R}_{>0}^{E}$, we call
\[
P_{I}\left(H\right)\coloneqq\conv\left(\left\{ \chi^{M}\mid M\subseteq E\text{ is a matching in }H\right\} \right)
\]
the integral hypergraph matching polytope of $H$ and 
\[
P\left(H\right)\coloneqq\left\{ x\in\mathbb{R}_{\ge0}^{E}\mid\sum_{e\in\delta\left(v\right)}x_{e}\le1\ \forall v\in V\right\} 
\]
the fractional hypergraph matching polytope of $H$.

The case where $k=2$ is the maximum weight matching problem in graphs,
which was extensively studied in the past. Here we focus on the results
related to the correlation gap of matchings. Guruganesh and Lee \cite{Guruganesh2018}
non-constructively prove that the correlation gap of the fractional
matching polytope is at least
\[
\mathbb{E}\left[\frac{1}{1+\Pois\left(2\right)}\right]=\frac{1-e^{-2}}{2}.
\]
In \cite{Bruggmann2019}, an explicit monotone $\frac{1-e^{-2}}{2}$-balanced
CRS is provided, thus giving a constructive proof of this lower bound
on the correlation gap. Moreover, \cite{Bruggmann2019} gives an optimal
monotone CRS for the bipartite matching polytope with balancedness
\[
\mathbb{E}\left[\frac{1}{1+\max\left\{ \Pois_{1}\left(1\right),\Pois_{2}\left(1\right)\right\} }\right].
\]
This scheme is then used to develop a monotone CRS with balancedness
strictly greater than $\frac{1-e^{-2}}{2}$ for general graphs, thus
proving that the original scheme for general graphs is not optimal
and that $\frac{1-e^{-2}}{2}$ is not a tight lower bound on the correlation
gap.

The RNC algorithm for general PIPs by Bansal et al.\,\cite{Bansal2012}
can be reformulated and applied to the hypergraph matching setting
(see Algorithm 8 in \cite{Brubach2019}). Given a hypergraph $H=\left(V,E\right)$
and a point $x$ in the standard LP relaxation of the hypergraph matching
polytope, the adapted algorithm constructs a matching where each hyperedge
$e\in E$ is included with probability at least $\frac{x_{e}}{\left|e\right|+1+o\left(\left|e\right|\right)}$,
where $\left|e\right|$ is the cardinality of the hyperedge (by Lemma
7 in \cite{Brubach2019}). Brubach et al.\,\cite{Brubach2019} improve
upon this algorithm and increase the inclusion probability to at least
$\frac{x_{e}}{\left|e\right|}\left(1-\exp\left(-\left|e\right|\right)\right)$
for every hyperedge $e\in E$ (see Algorithm 9 in \cite{Brubach2019}).
Moreover, the algorithm from \cite{Brubach2019} can be reformulated
as a combination of an independent sampling step and a $\frac{1-e^{-k}}{k}$-balanced
CRS for the natural LP relaxation of the hypergraph matching problem,
where $k\coloneqq\max\left\{ \left|e\right|\mid e\in E\right\} $.

From the point of view of approximation, everything revolves around
the following long-standing well-known conjecture.
\begin{conjecture}[\cite{Fueredi1993}, Conjecture 1.1]
\label{conj:furedi}For any hypergraph $H=\left(V,E\right)$ and
weight vector $w\in\mathbb{Z}_{\ge0}^{E}$, there exists a matching
$M\subseteq E$ such that
\[
\sum_{e\in M}\left(\left|e\right|-1+\frac{1}{\left|e\right|}\right)w_{e}\ge\max\left\{ \sum_{e\in H}w_{e}x_{e}\mid x\in P\left(H\right)\right\} .
\]
\end{conjecture}
Füredi at al.\,\cite{Fueredi1993} non-constructively show that Conjecture
\ref{conj:furedi} holds in the unweighted case (i.e.,\,if $w_{e}=1$
for every $e\in E$), for $k$-uniform hypergraphs, and for intersecting
hypergraphs (proved in \cite{Fueredi1993}, Theorems 1.2, 1.3, and
1.4 respectively). Chan and Lau \cite{Chan2011} give an algorithmic
proof of the conjecture for $k$-uniform hypergraphs. In \cite{Parekh2011,Parekh2015},
the conjecture is generalized to the $k$-uniform $b$-hypergraph
matching setting, and an optimal integrality gap of this family of
problems is obtained. The work of Bansal et al.\,\cite{Bansal2012}
gives $\left|e\right|+1$ instead of $\left|e\right|-1+\frac{1}{\left|e\right|}$
in Conjecture \ref{conj:furedi}, while Brubach et al.\,\cite{Brubach2019}
improve this to $\left|e\right|+O\left(\left|e\right|\cdot\exp\left(-\left|e\right|\right)\right)$
and slightly generalize the conjecture.

The (weighted) hypergraph matching problem is equivalent to the weighted
set packing problem: given a family of sets over a certain domain
and a weight assignment for each of the sets, find a subfamily of
pairwise-disjoint sets of maximum weight. A special case where all
sets in the family are of the same size $k$ and have weight $1$
is called the $k$-set packing problem. Therefore, the hardness result
for the $k$-set packing problem proved in \cite{Hazan2006}, which
we formulate below, shows that the more general hypergraph matching
problem is also NP-hard to approximate to within $\Omega\left(\frac{k}{\ln k}\right)$.
\begin{thm*}[\cite{Hazan2006}, Theorem 1.1]
It is NP-hard to approximate the $k$-set packing problem to within
$\Omega\left(\frac{k}{\log k}\right)$.
\end{thm*}

\subsection{\label{sec:intro klp}Knapsack problem}

This section focuses on the knapsack problem --- an integer program
of the following form:
\begin{align*}
\max\  & \sum_{i\in\left[n\right]}v_{i}x_{i}\\
\text{s.t.}\  & \sum_{i\in\left[n\right]}a_{i}x_{i}\le1,\\
 & x\in\left\{ 0,1\right\} ^{n},
\end{align*}
where $n\in\mathbb{Z}_{\ge1}$ is the number of items, $v\in\mathbb{R}_{>0}^{n}$
are item values and $a\in\left[0,1\right]^{n}$ are item sizes. Note
that the capacity of the knapsack can be set to $1$ by scaling the
sizes of all items. We call
\[
P_{I}\left(a\right)\coloneqq\conv\left(\left\{ \chi^{R}\mid\sum_{i\in R}a_{i}\le1,\ R\subseteq\left[n\right]\right\} \right)
\]
the integral knapsack polytope and
\[
P\left(a\right)\coloneqq\left\{ x\in\left[0,1\right]^{n}\mid\sum_{i\in\left[n\right]}a_{i}x_{i}\le1\right\} 
\]
the fractional knapsack polytope.

Checkuri et al.\,\cite{Chekuri2014} give a monotone $\left(b,1-2b\right)$-balanced
CRS for the fractional knapsack polytope. Moreover, it is shown that,
roughly speaking, one can get a $\left(1-\varepsilon,1-\varepsilon\right)$-balanced
CRS for a constant number ($m=O\left(1\right)$) of knapsack constraints
for any fixed $\varepsilon>0$ by guessing and enumeration tricks.
More specifically, Lemma 4.15 in \cite{Chekuri2014} provides a $\left(1-\varepsilon,1-\exp\left(-\Omega\left(\frac{\varepsilon^{2}}{\delta}\right)\right)\right)$-balanced
CRS that can be directly applied to instances with relatively small
item sizes (compared to the knapsack capacity), and Corollary 4.16
in \cite{Chekuri2014} shows how to apply this scheme to general instances
by using enumeration tricks. In addition, the former lemma allows
knapsack constraints to be combined with other types of constraints
in a black-box fashion. For more details, see Section 4.5 in \cite{Chekuri2014}.

Bansal et el.\,\cite{Bansal2012} present an $8k$-approximation
algorithm for $k$-CS-PIP based on the natural LP relaxation (discussed
in more detail in Sections \ref{sec:intro kcs} and \ref{sec:kcs}).
The algorithm consists of two steps: independent random sampling,
where a random set $R\subseteq\left[n\right]$ is generated by picking
each item $i\in\left[n\right]$ independently with probability $\frac{x_{j}}{4k}$,
and randomized alteration, where certain items are removed from $R$
to obtain a feasible set. As explained in Section \ref{sec:intro kcs},
the algorithm can thus be reformulated as a combination of an independent
sampling step and a $\frac{1}{8k}$-balanced CRS for the natural LP
relaxation of $k$-CS-PIP. In particular, by setting $k=1$, one obtains
a $\frac{1}{8}$-balanced CRS for the standard LP relaxation of the
knapsack problem, which is given explicitly in Algorithm \ref{alg:klp simple gen crs}.

\begin{algorithm2e}
\caption{\label{alg:klp simple gen crs}Contention resolution scheme for fractional knapsack polytope following from \cite{Bansal2012}.}
\KwIn{A size vector $a\in\left[0,1\right]^{n}$, a point $x\in\left[0,1\right]^{n}$, a set $R\subseteq\supp\left(x\right)$.}
\KwOut{A feasible solution $S\subseteq\left[n\right]$.}
Sample each item $j\in R$ independently with probability $\frac{1}{4}$ and let $S\subseteq R$ be the set of sampled items\;
For each item $j\in R$, mark it for deletion from $R$ if there exists an item $j'\in S\setminus\left\{j\right\}$ such that $a_{j'}>\frac{1}{2}$ or if $\sum_{j'\in S\colon a_{j'}\le\frac{1}{2}}a_{j'}>1$\;
Delete all marked items from $R$ and return the remaining items\;
\end{algorithm2e}

\subsection{\label{sec:intro kcs}$k$-column sparse packing integer problem}

In this section, we discuss $k$-column sparse packing integer programs,
which generalize the hypergraph matching and the knapsack problems.
A $k$-column sparse packing integer program ($k$-CS-PIP) with $n$
items and $m$ constraints is an integer program of the following
form:
\begin{align*}
\max\  & \sum_{j\in\left[n\right]}v_{j}x_{j}\\
\text{s.t.}\  & \sum_{j\in\left[n\right]}a_{ij}x_{j}\le1\quad\forall i\in\left[m\right],\\
 & x\in\left\{ 0,1\right\} ^{n},
\end{align*}
where $v\in\mathbb{R}_{>0}^{n}$ is the vector of item values, and
every column $a_{\cdot j}\in\left[0,1\right]^{m}$ of the constraint
matrix represents the sizes of the item $j\in\left[n\right]$ in each
of the $m$ knapsack constraints and has at most $k$ non-zero entries.
The natural LP relaxation of $k$-CS-PIP is
\begin{align*}
\max\  & \sum_{j\in\left[n\right]}v_{j}x_{j}\\
\text{s.t.}\  & \sum_{j\in\left[n\right]}a_{ij}x_{j}\le1\ \forall i\in\left[m\right],\\
 & x\in\left[0,1\right]^{n},
\end{align*}
and the strengthened LP relaxation (introduced in \cite{Bansal2012})
is
\begin{align*}
\max\  & \sum_{j\in\left[n\right]}v_{j}x_{j}\\
\text{s.t.}\  & \sum_{j\in\left[n\right]}a_{ij}x_{j}\le1\ \forall i\in\left[m\right],\\
 & \sum_{j\in\mathrm{big}\left(i\right)}x_{j}\le1\ \forall i\in\left[m\right],\\
 & x\in\left[0,1\right]^{n},
\end{align*}
where $\mathrm{big}\left(i\right)$ denotes the set of items $j\in\left[n\right]$
such that $a_{ij}>\frac{1}{2}$.

The $k$-CS-PIP problem is first considered in its full generality
by Pritchard \cite{Pritchard2009}, who also describes a $2^{k}k^{2}$-approximation
algorithm. In \cite{Pritchard2010}, an improved algorithm with approximation
ratio $2k^{2}+2$ is given.

Bansal et al.\,\cite{Bansal2012} give an $8k$-approximation algorithm
for $k$-CS-PIP based on the natural LP relaxation. The main idea
is to first independently sample and then perform a randomized alteration
step to obtain a feasible integral solution. As shown in the proof
of Theorem 3.3 in \cite{Bansal2012}, for every item $j\in\left[n\right]$,
the probability of it being returned is $\Pr\left[j\in R'\right]\ge\frac{x_{j}}{8k}$,
so the algorithm can be viewed as a combination of a randomized sampling
step (where each item $j\in\left[n\right]$ is sampled independently
with probability $x_{j}$) and a $\frac{1}{8k}$-balanced contention
resolution scheme, which we provide in Algorithm \ref{alg:kcs 8k approx}.
Furthermore, using the same general idea, Bansal et al. obtain an
$\left(ek+o\left(k\right)\right)$-approximation algorithm based on
the strengthened LP formulation and ultimately an $\left(\frac{e^{2}}{e-1}k+o\left(k\right)\right)$-approximation
algorithm for non-negative monotone submodular function maximization
over $k$-CS packing constraints.

\begin{algorithm2e}
\caption{\label{alg:kcs 8k approx}Contention resolution scheme for natural LP relaxation of $k$-CS-PIP from \cite{Bansal2012}.}
\KwIn{A $k$-column sparse constraint matrix $a\in\left[0,1\right]^{m\times n}$, a point $x\in\left[0,1\right]^{n}$ such that $\sum_{j=1}^{n}a_{ij} x_j \le 1$ for every $i\in\left[m\right]$, a set $R\subseteq\supp\left(x\right)$.}
\KwOut{A feasible solution $S\subseteq\left[n\right]$.}
Sample each item $j\in R$ independently with probability $\frac{1}{4k}$ and let $S\subseteq R$ be the set of sampled items\;
For each item $j\in S$, mark it for deletion from $S$ if there exists an item $j'\in S\setminus\left\{j\right\}$ and a constraint $i\in\left[m\right]$ such that $a_{ij}>0$ and $a_{ij'}>\frac{1}{2}$ or if there exists a constraint $i\in\left[m\right]$ such that $\sum_{j'\in S\colon a_{ij'}\le\frac{1}{2}}a_{ij'}>1$\;
Delete all marked items from $S$ and return the set of remaining items\;
\end{algorithm2e}

Using the algorithm from \cite{Bansal2012} as a base, Chekuri et
al.\,\cite{Chekuri2014} describe a $\left(b,1-2kb\right)$-balanced
CRS for the natural LP relaxation of $k$-CS-PIP and a $\left(b,1-k\left(2eb\right)^{W-1}\right)$-balanced
CRS for the case when the constraint matrix has width $W\ge2$.

Brubach et al.\,\cite{Brubach2019} provide a randomized rounding
algorithm based on the strengthened LP relaxation that has the approximation
ratio of at most $2k+\Theta\left(k^{0.8}\mathrm{poly}\log\left(k\right)\right)=2k+o\left(k\right)$
and can be broken down into an independent sampling step (where every
item $j\in\left[n\right]$ is independently sampled with probability
$x_{j}$) and a contentention resolution scheme. Using a theorem from
\cite{Chekuri2014}, the CRS is extended to submodular objectives
to obtain a randomized rounding algorithm with approximation ratio
$\frac{2k+o\left(k\right)}{\eta_{f}}$ for $k$-CS-PIP with non-negative
submodular objectives $f$. Here $\eta_{f}$ is the approximation
ratio for $\max\left\{ F\left(x\right)\mid x\in P_{\mathcal{I}}\cap\left\{ 0,1\right\} ^{n}\right\} $
where $F$ is the multilinear extension of $f$ and $P_{\mathcal{I}}$
is the $k$-CS-PIP polytope ($\eta_{f}=1-\frac{1}{e}$ if $f$ is
non-negative and monotone, and $\eta_{f}=0.385$ if $f$ is non-monotone).

Since the $k$-set packing problem is a special case of $k$-CS-PIP
with $a\in\left\{ 0,1\right\} ^{m\times n}$ (and at most $k$ non-zero
entries in each column), the hardness result for $k$-set packing
\cite{Hazan2006} also implies that it is NP-hard to approximate $k$-CS-PIP
to within $\Omega\left(\frac{k}{\log k}\right)$.

\subsection{\label{subsec:intro org}Our results}

In Section \ref{sec:cg ext prop}, we point out a technical issue
with the original definition of the correlation gap and resolve it
by first restricting the space of parameters (namely, requiring $v$
and $x$ to be strictly positive) and then continuously extending
the function to the largest possible region (see Lemma \ref{lem:cg cont}).
We note, however, that one may use the restricted parameter space
essentially without loss of generality (see remark after Corollary
\ref{cor:cg inf att}). In addition, we point out several useful properties
of the correlation gap, i.e.,\,its invariance with respect to scaling
of the value vector by a constant factor (Lemma \ref{lem:cg scale v}),
an upper bound involving the integrality gap (Lemma \ref{lem:cg ub by ig}),
and that the infimum over all $x$ for any $v$ is in fact attained
(Corollary \ref{cor:cg inf att}).

Section \ref{sec:hg} adapts the techniques from \cite{Guruganesh2018}
and \cite{Bruggmann2019} to the hypergraph matching setting. As a
result, we prove that the correlation gap of the fractional hypergraph
matching polytope is at least $\frac{1-e^{-k}}{k}$ (in Section \ref{subsec:hg non-construct})
and present an explicit monotone $\frac{1-e^{-k}}{k}$-balanced contention
resolution scheme for the hypergraph matching problem (Section \ref{subsec:hg crs}).
Thus, we improve upon the latest result of \cite{Brubach2019} (see
Algorithm 9) by gaining the monotonicity property while achieving
the same balancedness.

In Section \ref{sec:klp}, we consider the knapsack problem. Section
\ref{subsec:klp ck} focuses on the correlation gap of the fractional
knapsack polytope. We introduce the notion of a class-$k$ instance
of the knapsack problem where $k\in\mathbb{Z}_{\ge1}$ and show that
the correlation gap of any class-$k$ instance is at least $\frac{1-e^{-2}}{2}$
(Theorem \ref{thm:klp ck cg}). We conjecture that a family of example
class-$k$ instances we use as motivation (Example \ref{exa:klp ck tight cg})
attain the lower bound on the correlation gap; moreover, we believe
$\frac{1-e^{-2}}{2}$ is in fact a tight lower bound on the correlation
gap in general, i.e.,\,for any instance of the knapsack problem (Conjecture
\ref{conj:klp ck tightness}). As far as we know, this is the first
time the correlation gap of the fractional knapsack polytope is studied
directly (similar to how Guruganesh and Lee \cite{Guruganesh2018}
work with the correlation gap of matchings in graphs). In the next
two sections, we formulate and analyze several monotone contention
resolution schemes for the fractional knapsack polytope. Section \ref{subsec:klp big small}
gives an optimal monotone $\frac{1-e^{-2}}{2}$-balanced CRS for instances
where all item sizes are strictly bigger than $\frac{1}{2}$ and a
monotone $\frac{4}{9}$-balanced CRS for instances where all items
sizes are at most $\frac{1}{2}$. These two schemes show that the
correlation gap of class-$k$ knapsack instances is at least $\frac{1-e^{-2}}{2}$
(i.e.,\,together they form a constructive proof of Theorem \ref{thm:klp ck cg}).
Section \ref{subsec:klp gen} develops a monotone $0.279$-balanced
scheme for general case instances of the knapsack problem (with arbitrary
item sizes). This scheme is obtained by probabilistically combining
the previously mentioned monotone $\frac{1-e^{-2}}{2}$-balanced CRS
for class-$1$ instances and a simple monotone $\frac{1}{4}$-balanced
CRS for general instances, which is introduced in Section \ref{subsec:klp gen}.
Thus, we improve upon the monotone $\frac{1}{8}$-balanced contention
resolution scheme that follows from the work of Bansal et al.\,\cite{Bansal2012}
(see Sections \ref{sec:intro klp} and \ref{sec:intro kcs}). Nevertheless,
according to our conjecture about the correlation gap (Conjecture
\ref{conj:klp ck tightness}), the balancedness of our schemes in
the general case are far from optimal.

Section \ref{sec:kcs} is concerned with contention resolution for
$k$-CS-PIP. More specifically, we slightly modify the algorithm from
\cite{Brubach2019} to obtain a $\frac{1}{4k+o\left(k\right)}$-balanced
contention resolution scheme for the natural LP relaxation of $k$-CS-PIP.
In addition, we slightly refine the analysis (compared to \cite{Brubach2019})
and improve the $o\left(k\right)$ term. This scheme, when combined
with a sampling step where every item $j\in\left[n\right]$ is independently
sampled with probability $x_{j}$, immediately gives a $\left(4k+o\left(k\right)\right)$-approximation
algorithm for $k$-CS-PIP based on the natural LP relaxation. Nevertheless,
the question about the integrality gap of the natural LP relaxation
remains open, since the currently best known lower bound is $2k-1$,
attained by an example from \cite{Bansal2012} (which we provide in
Appendix \ref{subsec:ig kcs} for the sake of convenience).

In the appendix, Sections \ref{sec:tech hg}--\ref{sec:tech kcs}
contain various technical statements used in Sections \ref{sec:hg}--\ref{sec:kcs},
while Section \ref{sec:ig exas} provides examples attaining the integrality
gap of the three settings we consider, as well as for some modifications.

\subsection{\label{subsec:intro notation}Notation}

Throughout this work, we use the following notation.
\begin{itemize}
\item For $n\in\mathbb{Z}_{\ge1}$, we denote $\left[n\right]\coloneqq\left\{ 1,\dots,n\right\} $.
\item The $n$-dimensional all-ones vector is denoted as $1_{n}\in\mathbb{R}^{n}$.
\item For vectors $a,b\in\mathbb{R}^{n}$, the vector inequality $a\le b$
is interpreted coordinate-wise: $a_{i}\le b_{i}$ for every $i\in\left[n\right]$.
\item Given a ground set $E$ and a subset $M\subseteq E$, the characteristic
vector of $M$ is denoted as $\chi^{M}$.
\item Let $H=\left(V,E\right)$ be a hypergraph. For every $e\in E$, we
denote $N\left(e\right)\coloneqq\left\{ f\in E\mid f\cap e\neq\emptyset\right\} $
the neighborhood of the hyperedge $e$, including $e$ itself. For
every $v\in V$, the set of hyperedges incident to $v$ is denoted
as $\delta\left(v\right)\coloneqq\left\{ e\in E\mid v\in e\right\} $.
\item We use the following notation for random variables:
\begin{itemize}
\item $\Pois\left(\lambda\right)$ denotes a random variable drawn from
the Poisson distribution with parameter $\lambda>0$;
\item $\Bern\left(p\right)$ denotes a random variable indicating the result
of a Bernoulli trial with probability $p\in\left[0,1\right]$;
\item $\Binom\left(n,p\right)$ denotes a random variable drawn from the
binomial distribution with parameters $n\in\mathbb{Z}_{\ge0}$ and
$p\in\left[0,1\right]$;
\item $\Exp\left(\lambda\right)$ denotes a random variable drawn from the
exponential distribution with parameter $\lambda>0$.
\end{itemize}
\item We add a subscript to show independence of random variables: for example,
$\Pois_{1}\left(\xi\right)$ and $\Pois_{2}\left(\eta\right)$ are
independent (regardless of whether $\xi=\eta$).
\item The cumulative distribution function of a $\Binom\left(n,p\right)$
random variable is denoted as
\[
B\left(s;n,p\right)\coloneqq\sum_{j=0}^{s}\binom{n}{j}p^{j}\left(1-p\right)^{n-j}\quad\forall n\in\mathbb{Z}_{\ge1},\ p\in\left[0,1\right],\ s\in\left[n\right]\cup\left\{ 0\right\} ,
\]
while the cumulative distribution function of a $\Pois\left(\lambda\right)$
random variable is denoted as
\[
P\left(s;\lambda\right)\coloneqq\sum_{j=0}^{s}\frac{\lambda^{s}e^{-\lambda}}{s!}\quad\forall\lambda\in\mathbb{R}_{>0},\ s\in\mathbb{Z}_{\ge0}.
\]
\end{itemize}

\newpage{}

\section{\label{sec:cg ext prop}Extended definition and useful properties
of the correlation gap}

In this section, we resolve a technical issue with the original definition
of the correlation gap and point out several useful properties.

We assume the same setting as in Section \ref{sec:intro}. In what
follows below, let $E$ be a finite ground set, let $\mathcal{F}\subseteq2^{E}$
be a down-closed non-empty family of feasible sets, and let polytope
$P$ be a relaxation of $\mathcal{F}$. We consider the maximization
problem 
\begin{align*}
\max\  & \sum_{e\in S}v_{e}\\
\text{s.t.}\  & S\in\mathcal{F},
\end{align*}
where $v\in\mathbb{R}_{>0}^{E}$ is a vector of values, and its LP
relaxation
\begin{align*}
\max\  & \sum_{e\in E}v_{e}x_{e}\\
\text{s.t.}\  & x\in P.
\end{align*}

The definition of the correlation gap given in Definition \ref{def:intro cg orig}
is not entirely precise, as it is possible to have $\sum_{e\in E}v_{e}x_{e}=0$
in the denominator of the fraction, in which case the whole expression
is undefined. Below, we start by defining the correlation gap in a
smaller region where the technical issue does not arise and then continuously
extend it to the maximum possible space of parameters. In addition,
we define several variations of the correlation gap, since in future
sections we consider and compute not just the infimum over all values
of $x$ and $v$, but also the value of the fraction for fixed values
of $x$ and $v$ and the infimum over $x$ for a fixed $v$.
\begin{defn}
\label{def:cg ext}Let $x\in P\cap\mathbb{R}_{>0}^{E}$ and $v\in\mathbb{R}_{>0}^{E}$.
Then the correlation gap of $P$, $v$, and $x$ is defined as
\[
\CG\left(P,v,x\right)\coloneqq\frac{\mathbb{E}\left[\max\left\{ \sum\limits _{e\in S}v_{e}\mid S\subseteq R\left(x\right),\ S\in\mathcal{F}\right\} \right]}{\sum\limits _{e\in E}v_{e}x_{e}},
\]
where $R\left(x\right)\subseteq E$ is a random set that contains
every element $e\in E$ independently with probability $x_{e}$. Next,
let
\[
\CG\left(P,v\right)\coloneqq\inf\left\{ \CG\left(P,v,x\right)\mid x\in P\cap\mathbb{R}_{>0}^{E}\right\} 
\]
be the correlation gap of $P$ and $v$. Finally, we define the correlation
gap of $P$ as
\begin{align*}
\CG\left(P\right) & \coloneqq\inf\left\{ \CG\left(P,v\right)\mid v\in\mathbb{R}_{>0}^{E}\right\} \\
 & =\inf\left\{ \CG\left(P,v,x\right)\mid x\in P\cap\mathbb{R}_{>0}^{E},\ v\in\mathbb{R}_{>0}^{E}\right\} .
\end{align*}
To simplify notation, for any $R\subseteq E$, we denote the value
of an optimal integral solution supported on $R$ as
\[
\sigma_{v}\left(R\right)\coloneqq\max\left\{ \sum\limits _{e\in S}v_{e}\mid S\subseteq R,\ S\in\mathcal{F}\right\} .
\]
\end{defn}
\begin{rem*}
Since $\mathcal{F}=\left\{ S\subseteq E\mid\chi^{S}\in P\right\} $,
the correlation gaps $\CG\left(P,v,x\right)$, $\CG\left(P,v\right)$,
and $\CG\left(P\right)$ are well-defined, i.e.,\,it is not necessary
to specify $\mathcal{F}$ explicitly.
\end{rem*}
First, we note that the correlation gap is invariant with respect
to scaling of the value vector by a constant factor.
\begin{lem}
\label{lem:cg scale v}For any point $x\in P\cap\mathbb{R}_{>0}^{E}$,
values vector $v\in\mathbb{R}_{>0}^{E}$, and coefficient $\alpha\in\mathbb{R}_{>0}$,
we have 
\[
\CG\left(P,\alpha v,x\right)=\CG\left(P,v,x\right).
\]
\end{lem}
\begin{proof}
This property follows directly from the definition of $\CG\left(P,v,x\right)$.
\end{proof}
While the correlation gap is clearly upper bounded by $1$, the next
lemma gives a better bound.
\begin{lem}
\label{lem:cg ub by ig}For any $x\in P\cap\mathbb{R}_{>0}^{E}$ and
$v\in\mathbb{R}_{>0}^{E}$, let
\[
\mathrm{IG}\left(P\right)\coloneqq\max\left\{ \frac{\sum_{e\in R}v_{e}y_{e}}{\sigma_{v}\left(R\right)}\mid\supp\left(y\right)\subseteq R,\ y\in P;\ R\subseteq E\right\} 
\]
denote the worst-case integrality gap of the polytope $P$ with respect
to $\mathcal{F}$ (for $R=\emptyset$, we interpret the fraction inside
the maximum as $1$). Then
\[
\CG\left(P,v,x\right)\le\frac{1}{\mathrm{IG}\left(P\right)}.
\]
\end{lem}
\begin{proof}
By the definition of $\mathrm{IG}\left(P\right)$, for every $R\subseteq E$
we have
\[
\sigma_{v}\left(R\right)\le\frac{\sum_{e\in R}v_{e}}{\mathrm{IG}\left(P\right)}.
\]
Thus,
\[
\CG\left(P,v,x\right)\le\frac{1}{\mathrm{IG}\left(P\right)}\cdot\frac{\mathbb{E}\left[\sum\limits _{e\in R\left(x\right)}v_{e}\right]}{\sum\limits _{e\in E}v_{e}x_{e}}=\frac{1}{\mathrm{IG}\left(P\right)}.
\]
\end{proof}
Now we extend the definition of the correlation gap to a wider parameter
region while preserving continuity.
\begin{lem}
\label{lem:cg cont}The correlation gap $\CG\left(P,v,x\right)$ as
a function of $v$ and $x$ can be continuously extended to the region
\[
\mathcal{R}\coloneqq\left\{ \left(v,x\right)\in\mathbb{R}_{\ge0}^{E}\times P\mid\sum_{e\in E}v_{e}x_{e}\neq0\ \text{or}\ \supp\left(x\right)\in\mathcal{F}\right\} 
\]
and not outside of it, i.e.,\,it cannot be continuously extended
to any point $\left(v,x\right)\in\mathbb{R}_{\ge0}^{E}\times P$ such
that $\sum_{e\in E}v_{e}x_{e}=0$ and $\supp\left(x\right)\in\mathcal{F}$.
\end{lem}
\begin{proof}
We claim that the function 
\[
g\left(v,x\right)\coloneqq\begin{cases}
\frac{\mathbb{E}\left[\sigma_{v}\left(R\left(x\right)\right)\right]}{\sum\limits _{e\in E}v_{e}x_{e}}, & \sum_{e\in E}v_{e}x_{e}\neq0,\\
1, & \sum_{e\in E}v_{e}x_{e}=0\ \text{and}\ \supp\left(x\right)\in\mathcal{F}
\end{cases}
\]
is the continuous extension of $\CG\left(P,v,x\right)$ to the region
$\mathcal{R}$.

By construction, for any $\left(v,x\right)\in\mathbb{R}_{>0}^{E}\times\left(P\cap\mathbb{R}_{>0}^{E}\right)$
we have $g\left(v,x\right)=\CG\left(P,v,x\right)$, so $g\left(v,x\right)$
is an extension of $\CG\left(P,v,x\right)$. Thus, it only remains
to show that $g\left(v,x\right)$ is continuous in the region $\mathcal{R}$.

First, we show that $g\left(v,x\right)$ is continuous at every point
$\left(v,x\right)\in\mathcal{R}$ such that $\sum_{e\in E}v_{e}x_{e}\neq0$.
For any $R\subseteq E$, note that $\sigma_{v}\left(R\right)$ does
not depend on $x\in P$ and is a continuous function of $v\in\mathbb{R}_{\ge0}^{E}$
and that $\mathrm{Pr}\left[R\left(x\right)=R\right]$ does not depend
on $v\in\mathbb{R}_{\ge0}^{E}$ and is a continuous function of $x\in P$.
Therefore, 
\[
\mathbb{E}\left[\sigma_{v}\left(R\left(x\right)\right)\right]=\sum_{R\subseteq E}\sigma_{v}\left(R\right)\cdot\Pr\left[R\left(x\right)=R\right]
\]
is continuous at every point $\left(v,x\right)\in\mathbb{R}_{\ge0}^{E}\times P$.
Moreover, since $\sum_{e\in E}v_{e}x_{e}$ is continuous and (by our
assumption) non-zero, $g\left(v,x\right)$ is well-defined and continuous
for every $\left(v,x\right)\in\mathcal{R}$ such that $\sum_{e\in E}v_{e}x_{e}\neq0$
(as a composition of continuous functions).

Now, consider a point $\left(v,x\right)\in\mathcal{R}$ such that
$\sum_{e\in E}v_{e}x_{e}=0$ and $\supp\left(x\right)\in\mathcal{F}$.
We need to show that $\lim_{\bar{v}\to v,\bar{x}\to x}g\left(\bar{v},\bar{x}\right)=1=g\left(v,x\right)$.
Without loss of generality, we may assume that $\sum_{e\in E}\bar{v}_{e}\bar{x}_{e}\neq0$,
since otherwise $g\left(\bar{v},\bar{x}\right)=1$. For any subset
$R\subseteq E$, consider the following three cases.
\begin{enumerate}
\item Suppose that $\left|R\setminus\supp\left(x\right)\right|=0$. Then
\[
\Pr\left[R\left(x\right)=R\right]=\prod_{e\in R}\bar{x}_{e}\cdot\prod_{e\in\supp\left(x\right)\setminus R}\left(1-\bar{x}_{e}\right)\cdot\prod_{e\notin\supp\left(x\right)}\left(1-\bar{x}_{e}\right),
\]
\[
\sigma_{\bar{v}}\left(R\right)=\sum_{e\in R}\bar{v}_{e}.
\]
Therefore,
\begin{align*}
\sum_{R\subseteq E\colon\left|R\setminus\supp\left(x\right)\right|=0}\sigma_{\bar{v}}\left(R\right)\cdot\Pr\left[R\left(x\right)=R\right] & =\sum_{e\in\supp\left(x\right)}\bar{v}_{e}\bar{x}_{e}\cdot\prod_{e\notin\supp\left(x\right)}\left(1-\bar{x}_{e}\right)\\
 & \ge\sum_{e\in\supp\left(x\right)}\bar{v}_{e}\bar{x}_{e}\cdot\left(1-\sum_{e\notin\supp\left(x\right)}\bar{x}_{e}\right).
\end{align*}
\item Suppose that $\left|R\setminus\supp\left(x\right)\right|=1$. Let
$\left\{ h\right\} \coloneqq R\setminus\supp\left(x\right)$. Then
\[
\Pr\left[R\left(x\right)=R\right]=\bar{x}_{h}\cdot\prod_{e\in R\cap\supp\left(x\right)}\bar{x}_{e}\cdot\prod_{e\in\supp\left(x\right)\setminus R}\left(1-\bar{x}_{e}\right)\cdot\prod_{e\notin\supp\left(x\right)\cup\left\{ j\right\} }\left(1-\bar{x}_{e}\right),
\]
\[
\sigma_{\bar{v}}\left(R\right)\ge\bar{v}_{h}.
\]
Therefore,
\begin{align*}
\sum_{R\subseteq E\colon R\setminus\supp\left(x\right)=\left\{ h\right\} }\sigma_{\bar{v}}\left(R\right)\cdot\Pr\left[R\left(x\right)=R\right] & \ge\bar{v}_{h}\bar{x}_{h}\cdot\prod_{e\notin\supp\left(x\right)\cup\left\{ h\right\} }\left(1-\bar{x}_{e}\right)\\
 & \ge\bar{v}_{h}\bar{x}_{h}\cdot\left(1-\sum_{e\notin\supp\left(x\right)}\bar{x}_{e}\right).
\end{align*}
\item Suppose that $\left|R\setminus\supp\left(x\right)\right|\ge2$. Then
\[
\Pr\left[R\left(x\right)=R\right]=\prod_{e\in R\setminus\supp\left(x\right)}\bar{x}_{e}\cdot\prod_{e\in R\cap\supp\left(x\right)}\bar{x}_{e}\cdot\prod_{e\in\supp\left(x\right)\setminus R}\left(1-\bar{x}_{e}\right)\cdot\prod_{e\notin R\cup\supp\left(x\right)}\left(1-\bar{x}_{e}\right)\ge0,
\]
\[
\sigma_{\bar{v}}\left(R\right)\ge0.
\]
\end{enumerate}
Combining the results, we get:
\[
g\left(\bar{v},\bar{x}\right)=\frac{\mathbb{E}\left[\sigma_{\bar{v}}\left(R\left(\bar{x}\right)\right)\right]}{\sum\limits _{e\in E}\bar{v}_{e}\bar{x}_{e}}\ge\frac{\sum\limits _{e\in E}\bar{v}_{e}\bar{x}_{e}\cdot\left(1-\sum_{e\notin\supp\left(x\right)}\bar{x}_{e}\right)}{\sum\limits _{e\in E}\bar{v}_{e}\bar{x}_{e}}=1-\sum_{e\notin\supp\left(x\right)}\bar{x}_{e}.
\]
Since $\bar{x}_{e}\to0$ for every $e\notin\supp\left(x\right)$ as
$\bar{x}\to x$, we have $\lim_{\bar{v}\to v,\bar{x}\to x}g\left(\bar{v},\bar{x}\right)=1=g\left(v,x\right)$.
Thus, the function $g\left(v,x\right)$ is continuous at every point
$\left(v,x\right)\in\mathcal{R}$ such that $\sum_{e\in E}v_{e}x_{e}=0$
and $\supp\left(x\right)\in\mathcal{F}$.

Finally, consider any point $\left(v,x\right)\in\mathbb{R}_{\ge0}^{E}\times P$
such that $\sum_{e\in E}v_{e}x_{e}=0$ and $\supp\left(x\right)\notin\mathcal{F}$.
Let $h\in\supp\left(x\right)$, $\bar{x}\coloneqq x$, and $\tilde{x}\coloneqq x$.
For any parameter $\alpha\in\mathbb{R}_{>0}$, define 
\begin{align*}
\bar{v}_{e}\left(\alpha\right) & \coloneqq\begin{cases}
\alpha, & e=h,\\
v_{e}, & e\neq h,
\end{cases}\\
\tilde{v}_{e}\left(\alpha\right) & \coloneqq\begin{cases}
\alpha, & e\in\supp\left(x\right),\\
v_{e}, & e\notin\supp\left(x\right).
\end{cases}
\end{align*}
Then, on the one hand, for every $e\in E$ we have
\[
\lim_{\alpha\to+0}\bar{v}_{e}\left(\alpha\right)=\lim_{\alpha\to+0}\tilde{v}_{e}\left(\alpha\right)=v_{e},
\]
so $\left(\bar{v}\left(\alpha\right),\bar{x}\right)\to\left(v,x\right)$
and $\left(\tilde{v}\left(\alpha\right),\tilde{x}\right)\to\left(v,x\right)$
as $\alpha\to+0$. On the other hand, for any $\alpha\in\mathbb{R}_{>0}$
we have 
\[
\CG\left(P,\bar{v}\left(\alpha\right),\bar{x}\right)=\frac{\mathbb{E}\left[\sigma_{\bar{v}\left(\alpha\right)}\left(R\left(x\right)\right)\right]}{\sum_{e\in E}\bar{v}_{e}\left(\alpha\right)\bar{x}_{e}}=\frac{\alpha x_{h}}{\alpha x_{h}}=1,
\]
\[
\CG\left(P,\tilde{v}\left(\alpha\right),\tilde{x}\right)=\frac{\mathbb{E}\left[\sigma_{\tilde{v}\left(\alpha\right)}\left(R\left(x\right)\right)\right]}{\sum_{e\in E}\tilde{v}_{e}\left(\alpha\right)\tilde{x}_{e}}<\frac{\alpha\sum_{h\in\supp\left(x\right)}x_{h}}{\alpha\sum_{h\in\supp\left(x\right)}x_{h}}=1,
\]
where the inequality holds, since $\supp\left(x\right)\notin\mathcal{F}$.
Moreover, for any $\alpha\in\mathbb{R}_{>0}$ we have
\[
\CG\left(P,\bar{v}\left(\alpha\right),\bar{x}\right)=\CG\left(P,\bar{v}\left(1\right),\bar{x}\right),\quad\CG\left(P,\tilde{v}\left(\alpha\right),\tilde{x}\right)=\CG\left(P,\tilde{v}\left(1\right),\tilde{x}\right),
\]
i.e.,\,the two correlation gaps are constant as functions of $\alpha$
(see Remark above this lemma). Thus, 
\[
\lim_{\alpha\to+0}\CG\left(P,\bar{v}\left(\alpha\right),\bar{x}\right)\neq\lim_{\alpha\to+0}\CG\left(P,\bar{v}\left(\alpha\right),\bar{x}\right).
\]
Therefore, the function $\CG\left(P,v,x\right)$ cannot be continuously
extended to any point $\left(v,x\right)\notin\mathcal{R}$.
\end{proof}
Going forward, we denote the continuous extension of the correlation
gap defined in the proof of Lemma \ref{lem:cg cont} as $\CG\left(P,v,x\right)$,
slightly abusing the notation.
\begin{cor}
\label{cor:cg inf att}Since $P$ is a compact set, by the extreme
value theorem, for any $v\in\mathbb{R}_{>0}^{E}$ we have
\[
\CG\left(P,v\right)=\min\left\{ \CG\left(P,v,x\right)\mid x\in P\right\} .
\]
In other words, for any $v\in\mathbb{R}_{>0}^{E}$ there exists a
point $x^{*}\in P$ such that $\CG\left(P,v\right)=\CG\left(P,v,x^{*}\right)$.
\end{cor}
Finally, we note that one may essentially assume that $v$ and $x$
are strictly positive.
\begin{rem*}
Consider any point $\left(v,x\right)\in\mathcal{R}$ such that $\sum_{e\in E}v_{e}x_{e}\neq0$
and there exists $h\in E$ such that $v_{h}=0$ or $x_{h}=0$. Let
$E'\coloneqq\supp\left(v\right)\cup\supp\left(x\right)$ be the new
ground set (with all items $e\in E$ such that $v_{e}=0$ or $x_{e}=0$
removed), let $P'$, $v'$, and $x'$ be the projections of $P$,
$v$, and $x$ onto $\mathbb{R}^{E'}$, respectively (with components
corresponding to items $e\in E$ such that $v_{e}=0$ or $x_{e}=0$
ignored). Then $\CG\left(P,v,x\right)=\CG\left(P',v',x'\right)$.
In other words, for any $\left(v,x\right)\in\mathcal{R}$ such that
$\sum_{e\in E}v_{e}x_{e}\neq0$ the correlation gap (in its extended
definition) can be computed by first removing all items $e\in E$
such that $v_{e}=0$ or $x_{e}=0$ and then computing the correlation
gap of the pruned instance. Note that for any $\left(v,x\right)\in\mathbb{R}_{\ge0}^{E}\times P$
such that $\sum_{e\in E}v_{e}x_{e}=0$ this preprocessing step yields
an empty instance. In this case, it is reasonable to define the correlation
gap to be $1$, which is a neutral value in the following sense: since
we have $\CG\left(P,v,x\right)\le1$ for any $\left(v,x\right)\in\mathbb{R}_{\ge0}^{E}\times P$
such that $\sum_{e\in E}v_{e}x_{e}\neq0$, if we set $\CG\left(P,v,x\right)\coloneqq1$
for all other points $\left(v,x\right)\in\mathbb{R}_{\ge0}^{E}\times P$,
then the infima in the definitions of $\CG\left(P,v\right)$ and $\CG\left(P\right)$
are not affected. Moreover, note that if $\sum_{e\in E}v_{e}x_{e}=0$
and $v\neq0$, then $x$ is clearly not a maximizer of our objective
function over $P$, so this case never occurs in the relaxation and
rounding approach described in Section \ref{subsec:intro cg}, and
if $v=0$, then the original optimization problem is trivial. Keeping
these observations in mind, we assume that $v\in\mathbb{R}_{>0}^{E}$
and $x\in P\cap\mathbb{R}_{>0}^{E}$ going forward.
\end{rem*}

\newpage{}

\section{\label{sec:hg}Hypergraph matching problem}

\subsection{\label{subsec:hg non-construct}Non-constructive lower bound on the
correlation gap}

In this section, we provide a non-constructive proof of the lower
bound on the correlation gap of the fractional hypergraph matching
polytope (introduced in Section \ref{sec:intro hg}) by generalizing
and adapting the proof of Theorem 1.3 from \cite{Guruganesh2018}.

Let $H=\left(V,E\right)$ be a weighted hypergraph with $\left|e\right|\le k$
for every $e\in E$ and let $w\in\mathbb{R}_{>0}^{E}$ be the hyperedge
weight vector. Given $x\in\left[0,1\right]^{E}$, let $\mathcal{D}_{H,w,x}$
be the distribution over subhypergraphs of $H$ where each hyperedge
$e\in E$ is included independently with probability $x_{e}$ and
has weight $w_{e}$ if it is included. For any hypergraph $G$ sampled
from the distribution $\mathcal{D}_{H,w,x}$, let $\nu_{w}\left(G\right)$
be the weight of the maximum weight matching in $G$. 
\begin{thm}
\label{thm:hg nc}Let $x\in P\left(H\right)\cap\mathbb{R}_{>0}^{E}$.
Then
\[
\frac{\mathbb{E}_{G\sim\mathcal{D}_{H,w,x}}\left[\nu_{w}\left(G\right)\right]}{\sum_{e\in E}w_{e}x_{e}}\ge\frac{1-e^{-k}}{k}.
\]
\end{thm}
\begin{proof}
Define $F\colon\left[0,1\right]^{E}\to\mathbb{R}_{\ge0}$ as $F\left(x\right)\coloneqq\mathbb{E}_{G\sim\mathcal{D}_{H,w,x}}\left[\nu_{w}\left(G\right)\right]$.
Consider the function $\phi\left(t\right)\coloneqq F\left(tx\right)$
for $t\in\left[0,1\right]$. Then
\[
\frac{d\phi}{dt}=x\cdot\nabla F\left(tx\right)=\sum_{e\in E}x_{e}\left.\frac{\partial F}{\partial x_{e}}\right|_{tx}.
\]
For every $e\in E$,
\begin{align*}
\left.\frac{\partial F}{\partial x_{e}}\right|_{tx} & =\left.\frac{\partial\mathbb{E}_{G\sim\mathcal{D}_{H,w,x}}\left[\nu_{w}\left(G\right)\right]}{\partial x_{e}}\right|_{tx}\\
 & =\left.\frac{\partial\mathbb{E}_{G\sim\mathcal{D}_{H\setminus\left\{ e\right\} ,w,x}}\left[\nu_{w}\left(G\cup\left\{ e\right\} \right)\cdot x_{e}+\nu_{w}\left(G\setminus\left\{ e\right\} \right)\cdot\left(1-x_{e}\right)\right]}{\partial x_{e}}\right|_{tx}\\
 & =\mathbb{E}_{G\sim\mathcal{D}_{H,w,tx}}\left[\nu_{w}\left(G\cup\left\{ e\right\} \right)-\nu_{w}\left(G\setminus\left\{ e\right\} \right)\right].
\end{align*}
\begin{claim*}
In the setting of the theorem, for any hypergraph $G$ from the distribution
$\mathcal{D}_{H,w,x}$ we have 
\[
\sum_{e\in E}x_{e}\left(\nu_{w}\left(G\cup\left\{ e\right\} \right)-\nu_{w}\left(G\setminus\left\{ e\right\} \right)\right)+k\nu_{w}\left(G\right)\ge\sum_{e\in E}w_{e}x_{e}.
\]
\end{claim*}
\begin{proof}
Let $M\subseteq E\left(G\right)$ be a maximum weight matching in
$G$. Note:
\begin{align*}
 & \sum_{e\in E}x_{e}\left(\nu_{w}\left(G\cup\left\{ e\right\} \right)-\nu_{w}\left(G\setminus\left\{ e\right\} \right)\right)+k\nu_{w}\left(G\right)\\
\ge & \sum_{e\in E}x_{e}\left(\nu_{w}\left(G\cup\left\{ e\right\} \right)-\nu_{w}\left(G\right)\right)+k\sum_{f\in M}w_{f}\\
\ge & \sum_{e\in E}x_{e}\left(\nu_{w}\left(G\cup\left\{ e\right\} \right)-\nu_{w}\left(G\right)\right)+\sum_{f\in M}\left(w_{f}\sum_{e\in N\left(f\right)}x_{e}\right)\\
= & \sum_{e\in E}\left(\nu_{w}\left(G\cup\left\{ e\right\} \right)-\nu_{w}\left(G\right)+\sum_{f\in M\cap N\left(e\right)}w_{f}\right)x_{e}.
\end{align*}
The first inequality holds, since $\nu_{w}\left(G\right)\ge\nu_{w}\left(G\setminus\left\{ e\right\} \right)$
for every $e\in E$. The second inequality holds, since $x\in P\left(H\right)$
and $\left|f\right|\le k$ for every $f\in M\subseteq E$. For every
$e\in E$, let
\[
c_{e}\coloneqq\nu_{w}\left(G\cup\left\{ e\right\} \right)-\nu_{w}\left(G\right)+\sum_{f\in M\cap N\left(e\right)}w_{f}
\]
denote the coefficient of $x_{e}$ in the above sum. To prove the
claim, it suffices to show that $c_{e}\ge w_{e}$ for every $e\in E$.
Consider the following two cases.
\begin{enumerate}
\item If $M\cap N\left(e\right)=\emptyset$, then $\sum_{f\in M\cap N\left(e\right)}w_{f}=0$
and $M\cup\left\{ e\right\} $ is a matching. Since $M$ is a maximum
weight matching in $G$, it holds that $e\notin G$, so $\nu_{w}\left(G\cup\left\{ e\right\} \right)\ge\nu_{w}\left(G\right)+w_{e}$.
Thus, $c_{e}\ge w_{e}$.
\item Now suppose that $M\cap N\left(e\right)\neq\emptyset$.
\begin{enumerate}
\item If $\sum_{f\in M\cap N\left(e\right)}w_{f}\ge w_{e}$, then $c_{e}\ge\sum_{f\in M\cap N\left(e\right)}w_{f}\ge w_{e}$,
since $\nu_{w}\left(G\cup\left\{ e\right\} \right)\ge\nu_{w}\left(G\right)$.
\item If $\sum_{f\in M\cap N\left(e\right)}w_{f}<w_{e}$, then $M\cup\left\{ e\right\} \setminus N\left(e\right)$
is a matching of weight $\nu_{w}\left(G\right)+w_{e}-\sum_{f\in M\cap N\left(e\right)}w_{f}>\nu_{w}\left(G\right)$,
so $e\notin E\left(G\right)$. Therefore, $c_{e}\ge w_{e}$.
\end{enumerate}
\end{enumerate}
\end{proof}
By the claim,
\begin{align*}
\frac{d\phi}{dt} & =\sum_{e\in E}x_{e}\left.\frac{\partial F}{\partial x_{e}}\right|_{tx}\\
 & =\sum_{e\in E}\mathbb{E}_{G\sim\mathcal{D}_{H,w,tx}}\left[\nu_{w}\left(G\cup\left\{ e\right\} \right)-\nu_{w}\left(G\setminus\left\{ e\right\} \right)\right]\\
 & \ge\sum_{e\in E}w_{e}x_{e}-k\mathbb{E}_{G\sim\mathcal{D}_{H,w,tx}}\left[\nu_{w}\left(G\right)\right]\\
 & =\sum_{e\in E}w_{e}x_{e}-k\phi\left(t\right),
\end{align*}
which implies that
\[
\frac{d}{dt}\left(e^{kt}\phi\left(t\right)\right)=ke^{kt}\phi\left(t\right)+e^{kt}\frac{d\phi}{dt}\ge e^{kt}\sum_{e\in E}w_{e}x_{e}.
\]
Since $\phi\left(0\right)=0$, we have
\[
e^{k}\phi\left(1\right)\ge\sum_{e\in E}w_{e}x_{e}\int_{0}^{1}e^{kt}dt=\frac{e^{k}-1}{k}\sum_{e\in E}w_{e}x_{e},
\]
which proves the theorem.
\end{proof}
\begin{cor}
\label{cor:hg cg}The correlation gap of the fractional hypergraph
matching polytope is at least $\frac{1-e^{-k}}{k}$.
\end{cor}
\begin{rem*}
Note that by Lemma \ref{lem:cg ub by ig} applied to the hypergraph
$H$ from the projective plane example described in Appendix \ref{subsec:ig hg},
the correlation gap of the fractional hypergraph matching polytope
is at most $\frac{1}{k-1+\frac{1}{k}}$. This upper bound can be refined
by directly computing the correlation gap of this example: since any
integral solution contains at most one item, we have
\[
\sigma_{v}\left(R\right)=\begin{cases}
1, & R\neq\emptyset,\\
0, & R=\emptyset,
\end{cases}
\]
and thus 
\[
\CG\left(H,v,x\right)=\frac{1-\left(1-\frac{1}{k}\right)^{k^{2}-k+1}}{k-1+\frac{1}{k}}.
\]
\end{rem*}

\subsection{\label{subsec:hg crs}Contention resolution scheme}

In this section, we generalize the contention resolution scheme for
matchings in general (not necessarily bipartite) graphs presented
in \cite{Bruggmann2019} to the hypergraph setting (or, more specifically,
to the for the fractional hypergraph matching polytope, introduced
in Section \ref{sec:intro hg}).

Let $H=\left(V,E\right)$ be a hypergraph with $\left|e\right|\le k$
for every $e\in E$ for some $k\in\mathbb{Z}_{\ge1}$.
\begin{thm}
\label{thm:hg gen}There exists a monotone $\left(b,\frac{1-e^{-bk}}{bk}\right)$-balanced
contention resolution scheme for the fractional hypergraph matching
polytope.
\end{thm}
\begin{proof}
We claim that Algorithm \ref{alg:hg gen crs} is a monotone $\left(b,\frac{1-e^{-bk}}{bk}\right)$-balanced
CRS for the fractional hypergraph matching polytope.

\begin{algorithm2e}
\caption{\label{alg:hg gen crs}Contention resolution scheme for fractional hypergraph matching polytope.}
\KwIn{A hypergraph $H=\left(V,E\right)$, a point $x\in\left[0,1\right]^{E}$, a set $R\subseteq\supp\left(x\right)$.}
\KwOut{A point $y_{x}^{R}\in P_{I}\left(H\right)$ with $\supp\left(y_{x}^{R}\right)\subseteq R$.}
Sample each hyperedge $e\in R$ independently with probability $\frac{1-e^{-x_{e}}}{x_{e}}$ and let $\bar{R}$ be the set of sampled hyperedges\;
For each $e\in\bar{R}$, let $q_{e}\in\mathbb{Z}_{\ge1}$ be an independent realization of a $\Pois\left(x_{e}\right)$ random variable conditioned on it being at least $1$, and let $\left(y_{x}^{R}\right)_{e}\coloneqq\frac{q_{e}}{\sum_{f\in N\left(e\right)}q_{f}}$\;
For each $e\notin\bar{R}$, let $q_{e}\coloneqq0$ and let $\left(y_{x}^{R}\right)_{e}\coloneqq0$\;
Return $y_{x}^{R}$\;
\end{algorithm2e}

First, we prove that Algorithm \ref{alg:hg gen crs} is indeed a CRS
by showing that it always returns $y_{x}^{R}\in P_{I}\left(H\right)$.
Let $q\in\mathbb{Z}_{\ge0}^{E}$ be the vector defined in steps 2
and 3 of the algorithm. Define a random matching $M\subseteq\supp\left(q\right)$
in $H$ as follows: for every $e\in\supp\left(q\right)$, let $d_{e}$
be an independent realization of an $\Exp\left(q_{e}\right)$ random
variable; select $e$ only if $d_{e}<d_{f}$ for every $f\in N\left(e\right)\setminus\left\{ e\right\} $.
Clearly, $M$ is a matching. To simplify notation, let $\bar{q}_{e}\coloneqq\sum_{f\in N\left(e\right)\setminus\left\{ e\right\} }q_{f}$.
Then for every $e\notin\supp\left(q\right)$, we have $\mathbb{E}\left[\chi^{M}\left(e\right)\right]=\left(y_{x}^{R}\right)_{e}=0$,
and for every $e\in\supp\left(q\right)$, we have
\begin{align*}
\mathbb{E}\left[\chi^{M}\left(e\right)\right] & =\Pr\left[d_{e}<\min\left\{ d_{f}\mid f\in N\left(e\right)\setminus\left\{ e\right\} \right\} \right]\\
 & =\Pr\left[\Exp_{1}\left(q_{e}\right)<\Exp_{2}\left(\bar{q}_{e}\right)\right]\\
 & =\frac{q_{e}}{q_{e}+\bar{q}_{e}}=\frac{q_{e}}{\sum_{f\in N\left(e\right)}q_{f}}=\left(y_{x}^{R}\right)_{e}.
\end{align*}
The second equality follows from the two properties of exponential
random variables stated in Lemma \ref{lem:tech exp rvs}. Thus, $\mathbb{E}\left[\chi^{M}\right]=y_{x}^{R}$,
so $y_{x}^{R}\in P_{I}\left(H\right)$, which proves correctness of
Algorithm \ref{alg:hg gen crs}.

Next, we prove that our CRS is monotone by showing that $\mathbb{E}\left[\left(y_{x}^{A}\right)_{e}\right]\ge\mathbb{E}\left[\left(y_{x}^{B}\right)_{e}\right]$
for any $x\in P\left(H\right)$ and $e\in A\subseteq B\subseteq\supp\left(x\right)$.
Let $\bar{A}\left(x\right)\subseteq A$ and $\bar{B}\left(x\right)\subseteq B$
be the random subsets sampled in step 1 of Algorithm \ref{alg:hg gen crs}.
For every $f\in E$, let $q_{f}$ be an independent realization of
a $\Pois\left(x_{f}\right)$ random variable conditioned on it being
at least $1$. Then
\begin{align*}
\mathbb{E}\left[\left(y_{x}^{B}\right)_{e}\right] & =\frac{1-e^{-x_{e}}}{x_{e}}\cdot\sum_{\bar{B}\subseteq B\colon e\in\bar{B}}\mathbb{E}\left[\frac{q_{e}}{\sum_{f\in N\left(e\right)\cap\bar{B}}q_{f}}\mid\bar{B}\left(x\right)=\bar{B}\right]\\
 & \le\frac{1-e^{-x_{e}}}{x_{e}}\cdot\sum_{\bar{B}\subseteq B\colon e\in A\cap\bar{B}}\mathbb{E}\left[\frac{q_{e}}{\sum_{f\in N\left(e\right)\cap A\cap\bar{B}}q_{f}}\mid\bar{A}\left(x\right)=A\cap\bar{B}\right]\\
 & =\frac{1-e^{-x_{e}}}{x_{e}}\cdot\sum_{\bar{A}\subseteq A\colon e\in\bar{A}}\mathbb{E}\left[\frac{q_{e}}{\sum_{f\in N\left(e\right)\cap\bar{A}}q_{f}}\mid\bar{A}\left(x\right)=\bar{A}\right]\\
 & =\mathbb{E}\left[\left(y_{x}^{A}\right)_{e}\right].
\end{align*}
Thus, Algorithm \ref{alg:hg gen crs} is a monotone CRS.

\begin{algorithm2e}
\caption{\label{alg:hg gen merged}Merged sampling step with Algorithm \ref{alg:hg gen crs}.}
\KwIn{A hypergraph $H=\left(V,E\right)$, a point $x\in\left[0,1\right]^{E}$.}
\KwOut{A point $y_{x}\in P_{I}\left(H\right)$ with $\supp\left(y_{x}\right)\subseteq\supp\left(x\right)$.}
For each $e\in E$, let $q_{e}$ be an independent realization of a $\Pois\left(x_{e}\right)$ random variable\;
For each $e\in E$, let $y_{x}\left(e\right)\coloneqq\frac{q_{e}}{\sum_{f\in N\left(e\right)}q_{f}}$\;
Return $y_{x}$\;
\end{algorithm2e}

Finally, we prove that our CRS is $\left(b,\frac{1-e^{-bk}}{bk}\right)$-balanced.
To do this, we introduce auxiliary Algorithm \ref{alg:hg gen merged}
(which is not a CRS). Note that independently rounding a point $x\in\left[0,1\right]^{E}$
and then applying Algorithm \ref{alg:hg gen crs} is equivalent to
applying Algorithm \ref{alg:hg gen merged}. Indeed, we independently
round $x$ to $R\subseteq\supp\left(x\right)$ and then construct
$\bar{R}\subseteq R$ by independent subsampling with probability
$\frac{1-e^{-x_{e}}}{x_{e}}$ for every $e\in R$. Denoting $q_{e}$
an independent realization of a $\Pois\left(x_{e}\right)$ random
variable, we get:
\[
\Pr\left[e\in\bar{R}\right]=Pr\left[e\in\overline{R}\mid e\in R\right]\cdot\Pr\left[e\in R\right]=\frac{1-e^{-x_{e}}}{x_{e}}\cdot x_{e}=1-e^{-x_{e}}=\Pr\left[q_{e}\ge1\right].
\]
Therefore, the random variables $q_{e}$ defined in Algorithms \ref{alg:hg gen crs}
and \ref{alg:hg gen merged} have the same distribution, hence so
do the vectors $y_{x}^{R}$ and $y_{x}$ returned by the two algorithms.
We claim that for any $x\in bP\left(H\right)$ with $b\in\mathbb{R}_{\ge0}$
as the input, the output of Algorithm \ref{alg:hg gen merged} satisfies
\[
\mathbb{E}\left[\left(y_{x}\right)_{e}\right]\ge x_{e}\cdot\mathbb{E}\left[\frac{1}{1+\Pois\left(bk\right)}\right]=x_{e}\cdot\frac{1-e^{-bk}}{bk}.
\]
Using the properties of Poisson and Bernoulli random variables stated
in Lemma \ref{lem:tech pois rvs}, we obtain the following: 
\begin{align*}
\mathbb{E}\left[\left(y_{x}\right)_{e}\right] & =\mathbb{E}\left[\frac{\Pois_{e}\left(x_{e}\right)}{\sum_{f\in N\left(e\right)}\Pois_{f}\left(x_{f}\right)}\right]\\
 & =\mathbb{E}\left[\frac{\Pois_{e}\left(x_{e}\right)}{\Pois_{e}\left(x_{e}\right)+\sum_{f\in N\left(e\right)\setminus\left\{ e\right\} }\Pois_{f}\left(x_{f}\right)}\right]\\
 & =\mathbb{E}\left[\lim_{\ell\to\infty}\frac{\sum_{i=1}^{\ell}\Bern_{i}\left(\frac{x_{e}}{\ell}\right)}{\sum_{i=1}^{\ell}\Bern_{i}\left(\frac{x_{e}}{\ell}\right)+\Pois\left(\sum_{f\in N\left(e\right)\setminus\left\{ e\right\} }x_{f}\right)}\right]\\
 & =\lim_{\ell\to\infty}\sum_{i=1}^{\ell}\mathbb{E}\left[\frac{\Bern_{i}\left(\frac{x_{e}}{\ell}\right)}{\Bern_{i}\left(\frac{x_{e}}{\ell}\right)+\sum_{j\in\left[\ell\right]\setminus\left\{ i\right\} }\Bern_{j}\left(\frac{x_{e}}{\ell}\right)+\Pois\left(\sum_{f\in N\left(e\right)\setminus\left\{ e\right\} }x_{f}\right)}\right]\\
 & =\lim_{\ell\to\infty}\sum_{i=1}^{\ell}\frac{x_{e}}{\ell}\cdot\mathbb{E}\left[\frac{1}{1+\sum_{j\in\left[\ell\right]\setminus\left\{ i\right\} }\Bern_{j}\left(\frac{x_{e}}{\ell}\right)+\Pois\left(\sum_{f\in N\left(e\right)\setminus\left\{ e\right\} }x_{f}\right)}\right]\\
 & \ge\lim_{\ell\to\infty}\sum_{i=1}^{\ell}\frac{x_{e}}{\ell}\cdot\mathbb{E}\left[\frac{1}{1+\sum_{j\in\left[\ell\right]}\Bern_{j}\left(\frac{x_{e}}{\ell}\right)+\Pois\left(\sum_{f\in N\left(e\right)\setminus\left\{ e\right\} }x_{f}\right)}\right]\\
 & =x_{e}\cdot\mathbb{E}\left[\lim_{\ell\to\infty}\frac{1}{1+\sum_{j\in\left[\ell\right]}\Bern_{j}\left(\frac{x_{e}}{\ell}\right)+\Pois\left(\sum_{f\in N\left(e\right)\setminus\left\{ e\right\} }x_{f}\right)}\right]\\
 & =x_{e}\cdot\mathbb{E}\left[\frac{1}{1+\Pois_{e}\left(x_{e}\right)+\Pois\left(\sum_{f\in N\left(e\right)\setminus\left\{ e\right\} }x_{f}\right)}\right]\\
 & =x_{e}\cdot\mathbb{E}\left[\frac{1}{1+\Pois\left(\sum_{f\in N\left(e\right)}x_{f}\right)}\right].
\end{align*}
Since $x\in bP\left(H\right)$ and $\left|e\right|\le k$, we have
\[
\sum_{f\in N\left(e\right)}x_{f}=\sum_{v\in e}\sum_{f\in\delta\left(v\right)}x_{f}\le k\cdot b.
\]
Thus,
\begin{align*}
\mathbb{E}\left[\left(y_{x}\right)_{e}\right] & \ge x_{e}\cdot\mathbb{E}\left[\frac{1}{1+\Pois\left(bk\right)}\right]\\
 & =x_{e}\sum_{s=0}^{\infty}\frac{1}{s+1}\cdot\frac{\left(bk\right)^{s}e^{-bk}}{s!}\\
 & =\frac{x_{e}}{bk}\sum_{s=0}^{\infty}\frac{\left(bk\right)^{s+1}e^{-bk}}{\left(s+1\right)!}\\
 & =x_{e}\cdot\frac{1-e^{-bk}}{bk},
\end{align*}
which proves that Algorithm \ref{alg:hg gen crs} is a monotone $\left(b,\frac{1-e^{-bk}}{bk}\right)$-balanced
CRS.
\end{proof}
Since the monotone contention resolution scheme for matchings in general
graphs presented in \cite{Bruggmann2019} is non-optimal, we believe
that our contention resolution scheme presented in Algorithm \ref{alg:hg gen crs}
is not optimal and that the correlation gap of the fractional hypergraph
matching polytope is strictly greater than $\frac{1-e^{-k}}{k}$.

\newpage{}

\section{\label{sec:klp}Knapsack problem}

\subsection{\label{subsec:klp ck}Non-constructive lower bound on the correlation
gap of class-$k$ instances}

This section focuses on the family of class-$k$ instances of the
knapsack problem and shows that the correlation gap is at least $\frac{1-e^{-2}}{2}$,
which is conjectured to be the lower bound on the correlation gap
in general (i.e.,\,for any instance).

Throughout the section, we refer to a given pair of a size vector
$a\in\left[0,1\right]^{n}$ and a value vector $v\in\mathbb{R}_{>0}^{n}$
as an instance of the knapsack problem and write $\mathcal{I}\coloneqq\left(a,v\right)$.
For any $x\in P\left(a\right)$, we use the notation $\CG\left(a,v,x\right)$
as a shorthand for $\CG\left(P\left(a\right),v,x\right)$.

The notion of class-$k$ items and class-$k$ instances comes as a
natural generalization of the classifications of items by size used
in \cite{Bansal2012} and \cite{Brubach2019} (which distinguish ``big'',
``small'', ``medium'', and ``tiny'' items).
\begin{defn}
\label{def:klp ck classes}Let $a\in\left[0,1\right]^{n}$ be an item
size vector and let $k\in\mathbb{Z}_{\ge1}$. Then we say that $i\in\left[n\right]$
is a class-$k$ item if $a_{i}\in\left(\frac{1}{k+1},\frac{1}{k}\right]$,
and for any value vector $v\in\mathbb{R}_{>0}^{n}$, we call $\mathcal{I}=\left(a,v\right)$
a class-$k$ instance if $a_{i}\in\left(\frac{1}{k+1},\frac{1}{k}\right]$
for every $i\in\left[n\right]$ (i.e.,\,every item is a class-$k$
item).
\end{defn}
\begin{rem*}
Note that at most $k$ class-$k$ items can fit into the knapsack
simultaneously, so any integer solution of a class-$k$ instance consists
of at most $k$ items.
\end{rem*}
We begin with a motivating example.
\begin{example}
\label{exa:klp ck tight cg}For any $k,n\in\mathbb{Z}_{\ge1}$ such
that $n\ge k+1$ and any arbitrarily small $\varepsilon>0$, consider
the instance $\mathcal{I}_{k.n,\varepsilon}=\left(\frac{1+\varepsilon}{k+1}\cdot1_{n},1_{n}\right)$
and a point $x=\frac{k+1}{\left(1+\varepsilon\right)n}\cdot1_{n}\in P\left(a\right)$.
Then, by the Poisson limit theorem, we have
\begin{align*}
\lim_{\varepsilon\to+0,n\to+\infty}\CG\left(\frac{1+\varepsilon}{k+1}\cdot1_{n},1_{n},\frac{k+1}{\left(1+\varepsilon\right)n}\cdot1_{n}\right) & =\lim_{\varepsilon\to+0,n\to+\infty}\frac{1+\varepsilon}{k+1}\left(k-\sum_{j=0}^{k-1}B\left(j;n,\frac{k+1}{\left(1+\varepsilon\right)n}\right)\right)\\
 & =\frac{1}{k+1}\left(k-\sum_{j=0}^{k-1}P\left(j;k+1\right)\right).
\end{align*}
\end{example}
To simplify notation, we introduce the following function:
\[
F\left(s,\lambda\right)\coloneqq\frac{1}{\lambda}\left(s-\sum_{j=0}^{s-1}P\left(j;\lambda\right)\right)\quad\forall s\in\mathbb{Z}_{\ge1},\ \lambda\in\left[0,s+1\right].
\]
Thus, in the motivating example above we have 
\[
\lim_{\varepsilon\to+0,n\to+\infty}\CG\left(\frac{1+\varepsilon}{k+1}\cdot1_{n},1_{n},\frac{k+1}{\left(1+\varepsilon\right)n}\cdot1_{n}\right)=F\left(k,k+1\right).
\]

The main result of this section is the following theorem.
\begin{thm}
\label{thm:klp ck cg}For any class-$k$ instance $\mathcal{I}=\left(a,v\right)$
and any point $x\in P\left(a\right)$, we have $\CG\left(a,v,x\right)\ge\frac{1-e^{-2}}{2}.$
\end{thm}
\begin{proof}
Let $\mathcal{I}=\left(a,v\right)$ be a class-$k$ instance consisting
of $n$ items and let $x\in P\left(a\right)$. If $n\le k$, then
any subset $R\subseteq\left[n\right]$ is feasible, so $\CG\left(a,v,x\right)=1\ge\frac{1-e^{-2}}{2}$.
Thus, without loss of generality, we may assume that $n\ge k+1$.
In Lemmas \ref{lem:klp ck step 1}, \ref{lem:klp ck step 2}, and
\ref{lem:klp ck step 3} below, we prove a series of reductions that
allows us to derive the desired lower bound on the correlation gap
of class-$k$ instances. First, Lemma \ref{lem:klp ck step 1} shows
that replacing the value vector with the all-ones vector does not
increase the correlation gap. Next, we take the infimum over all $x\in P\left(a\right)$,
and by Lemma \ref{lem:klp ck step 2}, it is sufficient to consider
only points of the form $x=\mu\cdot1_{n}\in P\left(a\right)$ where
$\mu\in\left[0,1\right]$. Finally, by Lemma \ref{lem:klp ck step 3},
the correlation gap of instances with $v=1_{n}$ and $x=\mu\cdot1_{n}$
is lower bounded by $\frac{1-e^{-2}}{2}$. Thus, by combining the
resulting inequalities, we get:
\begin{align*}
\CG\left(a,v,x\right) & \ge\CG\left(a,1_{n},x\right)\\
 & \ge\inf\left\{ \CG\left(a,1_{n},\mu\cdot1_{n}\right)\mid\mu\in\left[0,1\right],\ \mu\cdot1_{n}\in P\left(a\right)\right\} \\
 & \ge\frac{1-e^{-2}}{2},
\end{align*}
which proves the theorem.
\end{proof}
\begin{rem*}
Note that Lemmas \ref{lem:klp ck step 1}, \ref{lem:klp ck step 2},
and \ref{lem:klp ck step 3} also hold for class-$k$ instances $\mathcal{I}=\left(a,v\right)$
with $n\le k$ items, since the correlation gap is $1$ regardless
of the values of $v$ and $x$.
\end{rem*}
\begin{lem}[Reduction step 1]
\label{lem:klp ck step 1}For any class-$k$ instance $\mathcal{I}=\left(a,v\right)$
with $n\ge k+1$ items and any point $x\in P\left(a\right)$, we have
\[
\CG\left(a,v,x\right)\ge\CG\left(a,1_{n},x\right).
\]
\end{lem}
\begin{proof}
Define
\[
v_{\text{avg}}\coloneqq\frac{\sum_{i\in\left[n\right]}v_{i}x_{i}}{\sum_{i\in\left[n\right]}x_{i}}.
\]
Without loss of generality, we may assume that
\[
v_{1}\ge\dots\ge v_{m}\ge v_{\text{avg}}\ge v_{m+1}\ge\dots\ge v_{n}
\]
for some $m\in\left[n-1\right]$. Consider a one-parameter family
of instances $\tilde{\mathcal{I}}\left(t\right)=\left(a,\tilde{v}\left(t\right)\right)$
with parameter $t\in\left[0,1\right]$ and value vector $\tilde{v}\left(t\right)\coloneqq\left(1-t\right)\cdot v+tv_{\text{avg}}\cdot1_{n}$.
Note the following.
\begin{enumerate}
\item The inequality $\tilde{v}_{i}\left(t\right)\ge\tilde{v}_{j}\left(t\right)$
holds if and only if $v_{i}\ge v_{j}$.
\item Since $\mathcal{I}\left(t\right)$ is a class-$k$ instance, for any
$t\in\left[0,1\right]$ and any sampled set $R\subseteq\left[n\right]$,
the item $i$ is a part of the optimal integral solution if and only
if $\left|R\cap\left[i-1\right]\right|\le k-1$.
\item By construction of $v_{\text{avg}}$ and $\tilde{v}\left(t\right)$,
we have
\[
\sum_{i\in\left[n\right]}\tilde{v}_{i}\left(t\right)x_{i}=\left(1-t\right)\sum_{i\in\left[n\right]}v_{i}x_{i}+v_{\text{avg}}t\sum_{i\in\left[n\right]}x_{i}=\sum_{i\in\left[n\right]}v_{i}x_{i}.
\]
\end{enumerate}
Putting the above observations together, we get:
\[
\CG\left(a,v,x\right)-\CG\left(a,\tilde{v}\left(t\right),x\right)=t\cdot\frac{\sum_{i\in\left[n\right]}\left(v_{i}-v_{\text{avg}}\right)x_{i}\cdot\Pr\left[\left|R\cap\left[i-1\right]\right|\le k-1\right]}{\sum_{i\in\left[n\right]}v_{i}x_{i}}.
\]
Note that $\Pr\left[\left|R\cap\left[i\right]\right|\le k-1\right]\le\Pr\left[\left|R\cap\left[i-1\right]\right|\le k-1\right]$
for every $i\in\left[n\right]$, and we have $v_{i}-v_{\text{avg}}\ge0$
for every $i\in\left[m\right]$ and $v_{\text{avg}}-v_{i}\ge0$ for
every $i\in\left\{ m+1,\dots,n\right\} $. Thus,
\begin{align*}
\sum_{i\in\left[n\right]}\left(v_{i}-v_{\text{avg}}\right)x_{i}\cdot\Pr\left[\left|R\cap\left[i-1\right]\right|\le k-1\right] & =\sum_{i=1}^{m}\left(v_{i}-v_{\text{avg}}\right)x_{i}\cdot\Pr\left[\left|R\cap\left[i-1\right]\right|\le k-1\right]\\
 & -\sum_{i=m+1}^{n}\left(v_{\text{avg}}-v_{i}\right)x_{i}\cdot\Pr\left[\left|R\cap\left[i-1\right]\right|\le k-1\right]\\
 & \ge\sum_{i=1}^{m}\left(v_{i}-v_{\text{avg}}\right)x_{i}\cdot\Pr\left[\left|R\cap\left[m-1\right]\right|\le k-1\right]\\
 & -\sum_{i=m+1}^{n}\left(v_{\text{avg}}-v_{i}\right)x_{i}\cdot\Pr\left[\left|R\cap\left[m-1\right]\right|\le k-1\right]\\
 & =\sum_{i\in\left[n\right]}\left(v_{i}-v_{\text{avg}}\right)x_{i}\cdot\Pr\left[\left|R\cap\left[m-1\right]\right|\le k-1\right]\\
 & =0.
\end{align*}
As a result, we have $\CG\left(a,v,x\right)\ge\CG\left(a,\tilde{v}\left(t\right),x\right)$
for every $t\in\left[0,1\right]$. In particular, for $t=1$ we get
$\CG\left(a,v,x\right)\ge\CG\left(a,v_{\text{avg}}\cdot1_{n},x\right)$.
Since by Lemma \ref{lem:cg scale v} scaling the value vector by a
non-negative factor does not change the correlation gap, we have $\CG\left(a,v_{\text{avg}}\cdot1_{n},x\right)=\CG\left(a,1_{n},x\right)$,
so the lemma holds.
\end{proof}
\begin{lem}[Reduction step 2]
\label{lem:klp ck step 2}For any class-$k$ instance $\mathcal{I}=\left(a,1_{n}\right)$
with $n\ge k+1$ items, we have
\begin{equation}
\inf\left\{ \CG\left(a,1_{n},\tilde{x}\right)\mid\tilde{x}\in P\left(a\right)\right\} =\inf\left\{ \CG\left(a,1_{n},\mu\cdot1_{n}\right)\mid\mu\in\left[0,1\right],\ \mu\cdot1_{n}\in P\left(a\right)\right\} .\label{eq:klp ck step 2 claim}
\end{equation}
\end{lem}
\begin{proof}
By Corollary \ref{cor:cg inf att}, there exists a point $x\in P\left(a\right)$
attaining the minimum on the left hand side of (\ref{eq:klp ck step 2 claim}).
Suppose that $x_{i}\neq x_{j}$ for some $i\neq j\in\left[n\right]$.
Since items can be reordered arbitrarily, without loss of generality,
we may assume that $x_{1}\neq x_{2}$.

First, consider the special case where $\left|\left\{ i\in\left\{ 3,\dots,n\right\} \mid x_{i}=1\right\} \right|\ge k$.
Since $x\in P\left(a\right)$ and $a_{i}>\frac{1}{k+1}$ for every
$i\in\left[n\right]$, we have
\[
1\ge\sum_{i\in\left[n\right]}a_{i}x_{i}>\frac{1}{k+1}\cdot\left|\left\{ i\in\left\{ 3,\dots,n\right\} \mid x_{i}=1\right\} \right|,
\]
thus $\left|\left\{ i\in\left\{ 3,\dots,n\right\} \mid x_{i}=1\right\} \right|=k$.
Again, since $x\in P\left(a\right)$ and $a_{i}>\frac{1}{k+1}$ for
every $i\in\left[n\right]$, we have
\[
1\ge\sum_{i\in\left[n\right]}a_{i}x_{i}>\frac{x_{1}+x_{2}+k}{k+1},
\]
so $x_{1}+x_{2}<1$. Let $j\in\left\{ 3,\dots,n\right\} $ be an item
such that $x_{j}=1$. Thus, $x\in P\left(a\right)$ attains the infimum
on the left hand side of (\ref{eq:klp ck step 2 claim}) and satisfies
$x_{1}\neq x_{j}$ and $\left|\left\{ i\in\left[n\right]\setminus\left\{ 1,j\right\} \mid x_{i}=1\right\} \right|=k-1<k$
(since $x_{2}\neq1$). Therefore, without loss of generality, we may
assume that $\left|\left\{ i\in\left\{ 3,\dots,n\right\} \mid x_{i}=1\right\} \right|<k$.

Now, let $R\subseteq\left[n\right]$ be a random subset where each
item $i\in\left[n\right]$ is included independently with probability
$x_{i}$. To simplify notation, we denote $R'\coloneqq R\setminus\left\{ 1,2\right\} $.
Then
\begin{align}
\sum_{t=0}^{n}\Pr\left[\left|R\right|=t\right]\cdot\min\left\{ k,t\right\}  & =\sum_{s=0}^{n-2}\Pr\left[\left|R'\right|=s\right]\cdot\left[x_{1}x_{2}\min\left\{ k,s+2\right\} \right.\nonumber \\
 & +\left(x_{1}\left(1-x_{2}\right)+x_{2}\left(1-x_{1}\right)\right)\min\left\{ k,s+1\right\} \nonumber \\
 & \left.+\left(1-x_{1}\right)\left(1-x_{2}\right)\min\left\{ k,s\right\} \right]\nonumber \\
 & =\sum_{s=0}^{n-2}\Pr\left[\left|R'\right|=s\right]\cdot\min\left\{ k,s\right\} \nonumber \\
 & +\Pr\left[\left|R'\right|\le k-2\right]\cdot\left(2x_{1}x_{2}+x_{1}\left(1-x_{2}\right)+x_{2}\left(1-x_{1}\right)\right)\nonumber \\
 & +\Pr\left[\left|R'\right|=k-1\right]\cdot\left(x_{1}x_{2}+x_{1}\left(1-x_{2}\right)+x_{2}\left(1-x_{1}\right)\right)\nonumber \\
 & =\sum_{s=0}^{n-2}\Pr\left[\left|R'\right|=s\right]\cdot\min\left\{ k,s\right\} \nonumber \\
 & +\Pr\left[\left|R'\right|\le k-2\right]\cdot\left(x_{1}+x_{2}\right)\nonumber \\
 & -\Pr\left[\left|R'\right|=k-1\right]\cdot x_{1}x_{2}.\label{eq:klp ck step 2 cg}
\end{align}
Consider the point $x'\in P\left(a\right)$ defined as follows: 
\[
x_{i}'=\begin{cases}
\frac{x_{1}+x_{2}}{2}, & i\in\left\{ 1,2\right\} ,\\
x_{i}, & i\in\left\{ 3,\dots,n\right\} .
\end{cases}
\]
Since $\sum_{i\in\left[n\right]}x_{i}'=\sum_{i\in\left[n\right]}x_{i}$
and $x_{1}'+x_{2}'=x_{1}+x_{2}$, from (\ref{eq:klp ck step 2 cg})
we get:
\[
\CG\left(a,1_{n},x\right)-\CG\left(a,1_{n},x'\right)=\frac{\left(x_{1}'x_{2}'-x_{1}x_{2}\right)\cdot\Pr\left[\left|R'\right|=k-1\right]}{\sum_{i\in\left[n\right]}x_{i}}.
\]
By construction of $x'$, we have
\[
x_{1}'x_{2}'-x_{1}x_{2}=\left(\frac{x_{1}+x_{2}}{2}\right)^{2}-x_{1}x_{2}=\left(\frac{x_{1}-x_{2}}{2}\right)^{2}>0.
\]
Moreover, we have $\Pr\left[\left|R'\right|=k-1\right]>0$, since
$n>k$ and $\left|\left\{ i\in\left\{ 3,\dots,n\right\} \mid x_{i}=1\right\} \right|<k$.
As a result, we get $\CG\left(a,1_{n},x\right)>\CG\left(a,1_{n},x'\right)$,
which contradicts the optimality of $x$ and proves that the infimum
on the left hand side of (\ref{eq:klp ck step 2 claim}) is attained
by some $x=\mu\cdot1_{n}$ with $\mu\in\left[0,1\right]$.
\end{proof}
\begin{lem}[Reduction step 3]
\label{lem:klp ck step 3}For any class-$k$ instance $\mathcal{I}=\left(a,1_{n}\right)$
with $n\ge k+1$ items and any $\mu\in\left[0,1\right]$ such that
$\mu\cdot1_{n}\in P\left(a\right)$ we have 
\[
\CG\left(a,1_{n},\mu\cdot1_{n}\right)\ge\frac{1-e^{-2}}{2}.
\]
\end{lem}
\begin{proof}
Define $\lambda\coloneqq\mu n$. Since $\mu\cdot1_{n}\in P\left(a\right)$,
we have
\[
1\ge\sum_{i\in\left[n\right]}a_{i}\cdot\frac{\lambda}{n}\ge n\cdot\frac{1}{k+1}\cdot\frac{\lambda}{n},
\]
so $\lambda\in\left[0,k+1\right]$. Using the cumulative distribution
function of the binomial distribution, we can write 
\begin{align*}
\CG\left(a,1_{n},\frac{\lambda}{n}\cdot1_{n}\right) & =\frac{1}{\lambda}\sum_{s=0}^{n}\binom{n}{s}\left(\frac{\lambda}{n}\right)^{s}\left(1-\frac{\lambda}{n}\right)^{n-s}\cdot\min\left\{ s,k\right\} \\
 & =\frac{1}{\lambda}\left(k-\sum_{s=0}^{k-1}B\left(s;n,\frac{\lambda}{n}\right)\right).
\end{align*}
Note the similarity to Example \ref{exa:klp ck tight cg} we introduced
in the beginning of this section. Consider the following three cases.
\begin{enumerate}
\item Suppose that $\lambda\in\left[0,1\right]$. Then
\begin{align*}
\CG\left(a,1_{n},\frac{\lambda}{n}\cdot1_{n}\right) & \ge\frac{1}{\lambda}\left(k-B\left(0;n,\frac{\lambda}{n}\right)-\sum_{s=1}^{k-1}1\right)\\
 & =\frac{1-\left(1-\frac{\lambda}{n}\right)^{n}}{\lambda}\\
 & \ge\frac{1-e^{-\lambda}}{\lambda}\\
 & \ge\frac{1-e^{-2}}{2}.
\end{align*}
The first inequality holds since $B\left(s;n,\frac{\lambda}{n}\right)\le1$
for any $n\in\mathbb{Z}_{\ge1}$ and $s\in\left[n\right]\cup\left\{ 0\right\} $.
The second inequality holds since $e^{x}\ge1+x$ for any $x\in\mathbb{R}$.
The last inequality holds since $x\mapsto\frac{1-e^{-x}}{x}$ is a
monotonically decreasing function of $x\in\mathbb{R}_{>0}$.
\item Suppose that $\lambda\in\left(1,k\right)$. Then
\begin{align*}
\CG\left(a,1_{n},\frac{\lambda}{n}\cdot1_{n}\right) & \ge\frac{1}{\lambda}\left(k-\sum_{s=0}^{\left\lfloor \lambda\right\rfloor -1}P\left(\lambda,s\right)-\sum_{s=\left\lfloor \lambda\right\rfloor }^{k-1}1\right)\\
 & =\frac{1}{\lambda}\left(\left\lfloor \lambda\right\rfloor -\sum_{s=0}^{\left\lfloor \lambda\right\rfloor -1}P\left(\lambda,s\right)\right)\\
 & =F\left(\left\lfloor \lambda\right\rfloor ,\lambda\right)\\
 & \ge F\left(\left\lfloor \lambda\right\rfloor ,\left\lfloor \lambda\right\rfloor +1\right)\\
 & \ge\frac{1-e^{-2}}{2}.
\end{align*}
The first inequality holds, since $B\left(s;n,\frac{\lambda}{n}\right)\le P\left(s;\lambda\right)$
for any $s\in\left[\left\lfloor \lambda\right\rfloor -1\right]\cup\left\{ 0\right\} $
by Lemma \ref{lem:tech klp cdf} and since $B\left(s;n,\frac{\lambda}{n}\right)\le1$
for any $s\in\left\{ \left\lfloor \lambda\right\rfloor ,\dots,k-1\right\} $.
The second and the third inequality hold by Lemmas \ref{lem:tech klp f dec}
and \ref{lem:tech klp f lb}, respectively.
\item Suppose that $\lambda\in\left[k,k+1\right]$. Then
\begin{align*}
\CG\left(a,1_{n},\frac{\lambda}{n}\cdot1_{n}\right) & \ge\frac{1}{\lambda}\left(k-\sum_{s=0}^{k-1}P\left(\lambda,s\right)\right)\\
 & =F\left(k,\lambda\right)\\
 & \ge F\left(k,k+1\right)\\
 & \ge\frac{1-e^{-2}}{2}.
\end{align*}
The first inequality holds, since by Lemma \ref{lem:tech klp cdf},
we have $B\left(s;n,\frac{\lambda}{n}\right)\le P\left(s;\lambda\right)$
for any $s\in\left[k-1\right]\cup\left\{ 0\right\} $. The second
and the third inequality hold by Lemmas \ref{lem:tech klp f dec}
and \ref{lem:tech klp f lb}, respectively.
\end{enumerate}
Thus, $\CG\left(a,1_{n},\frac{\lambda}{n}\cdot1_{n}\right)\ge\frac{1-e^{-2}}{2}$
holds for any $\lambda\in\left[0,k+1\right]$.
\end{proof}
\begin{figure}
\caption{\label{fig:klp simp}Numerical evidence for Conjecture \ref{conj:klp simp}.}

\noindent \centering{}\subfloat[\label{fig:klp simp gf}Evidence for (\ref{eq:klp simp gf 1}) and
(\ref{eq:klp simp gf 2}).]{
\noindent \centering{}\includegraphics[width=0.475\columnwidth,type=eps,natwidth=631,natheight=289]{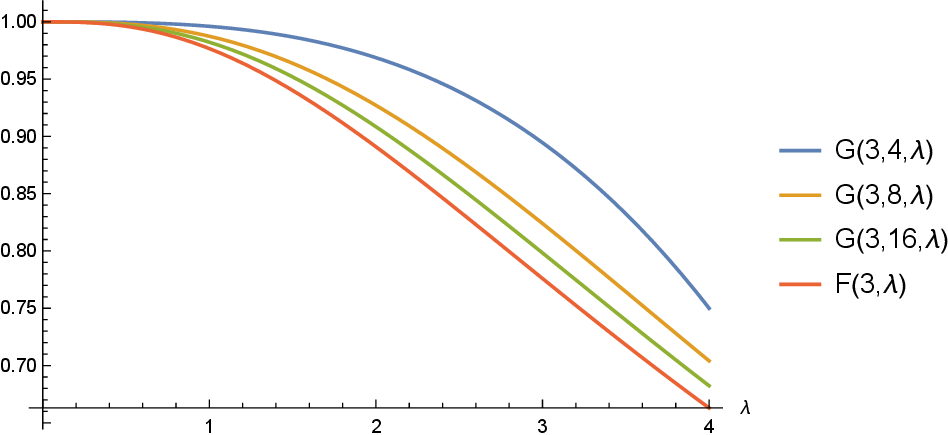}}\hfill{}\subfloat[\label{fig:klp simp ff}Evidence for (\ref{eq:klp simp ff 1}) and
(\ref{eq:klp simp ff 2}).]{
\noindent \centering{}\includegraphics[width=0.475\columnwidth,type=eps,natwidth=450,natheight=210]{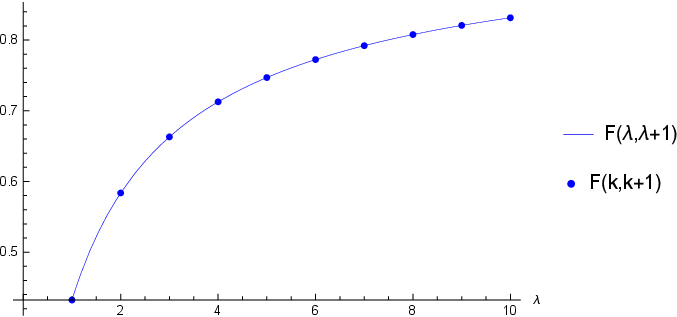}}
\end{figure}

Now that the proof of Theorem \ref{thm:klp ck cg} is complete, we
would like to explain a conjectured strengthening of our result. We
start with the following technical conjecture.
\begin{conjecture}
\label{conj:klp simp}For any $n\in\mathbb{Z}_{\ge1}$, $k\in\mathbb{Z}_{\ge1}$
such that $n\ge k+1$, and $\lambda\in\left[0,k+1\right]$, let
\[
G\left(k,n,\lambda\right)\coloneqq\frac{1}{\lambda}\left(k-\sum_{j=0}^{k-1}B\left(n,\frac{\lambda}{n},j\right)\right).
\]
Then the following inequalities hold:
\begin{equation}
G\left(k,n_{1},\lambda\right)\ge G\left(k,n_{2},\lambda\right)\ge F\left(k,\lambda\right)\quad\forall k+1\le n_{1}\le n_{2},\ n_{1},n_{2}\in\mathbb{Z}_{\ge1},\label{eq:klp simp gf 1}
\end{equation}
\begin{equation}
G\left(k,n,\lambda_{1}\right)\ge G\left(k,n,\lambda_{2}\right)\quad\forall0\le\lambda_{1}\le\lambda_{2}\le k+1,\ \lambda_{1},\lambda_{2}\in\mathbb{R}_{\ge0},\label{eq:klp simp gf 2}
\end{equation}
\begin{equation}
F\left(k_{2},k_{2}+1\right)\ge F\left(k_{1},k_{1}+1\right)\quad\forall k_{1}\le k_{2},\ k_{1},k_{2}\in\mathbb{Z}_{\ge1}.\label{eq:klp simp ff 1}
\end{equation}
Moreover, if we extend the function $F$ continuously to $k\in\mathbb{R}_{\ge1}$
by replacing $\left(k-1\right)!$ with $\Gamma\left(k\right)$ in
the expression obtained in Lemma \ref{lem:tech klp f dec}, i.e.,\,defining
\[
F\left(k,\lambda\right)\coloneqq\frac{1}{\lambda}\left(k+\left(\lambda-k\right)\frac{\Gamma\left(k,\lambda\right)}{\Gamma\left(k\right)}-\frac{\lambda^{k}e^{-\lambda}}{\Gamma\left(k\right)}\right)
\]
for every $k\in\mathbb{R}_{\ge1}$ and $\lambda\in\left[0,k+1\right]$,
then the following strengthening of the inequality (\ref{eq:klp simp ff 1})
holds:
\begin{equation}
F\left(\lambda_{2},\lambda_{2}+1\right)\ge F\left(\lambda_{1},\lambda_{1}+1\right)\quad\forall\lambda_{1}\le\lambda_{2},\ \lambda_{1},\lambda_{2}\in\mathbb{R}_{\ge1}.\label{eq:klp simp ff 2}
\end{equation}
\end{conjecture}
This conjecture seems to hold numerically, as shown in Figure \ref{fig:klp simp}.
Moreover, Lemma \ref{lem:tech klp cdf} implies that (\ref{eq:klp simp gf 1})
holds for any $\lambda\in\left[k,k+1\right]$, but it is not sufficient
to prove the inequality for all $\lambda\in\left[0,k+1\right]$.

Conjecture \ref{conj:klp simp} leads to several potential simplifications
and strengthened results. Indeed, from Lemma \ref{lem:klp ck step 3}
and Conjecture \ref{conj:klp simp} we know that $\CG\left(a,1_{n},\frac{\lambda}{n}\cdot1_{n}\right)=G\left(k,n,\lambda\right)$,
so the inequality (\ref{eq:klp simp gf 1}) together with Lemma \ref{lem:tech klp f dec}
would imply that $\CG\left(a,1_{n},\frac{\lambda}{n}\cdot1_{n}\right)\ge F\left(k,k+1\right)$.
By the conjectured inequality (\ref{eq:klp simp ff 1}), this bound
would be stronger than the one we obtained in Lemma \ref{lem:klp ck step 3}.
With this improved bound, our series of reductions would imply that
for any class-$k$ instance $\mathcal{I}=\left(a,v\right)$ and any
point $x\in P\left(a\right)$, we have $\CG\left(a,v,x\right)\ge F\left(k,k+1\right)$.
This leads us to believe that, roughly speaking, class-$1$ items
are the bottleneck for the correlation gap. Thus, we believe that
the lower bound on the correlation gap of class-$k$ instances proved
in Theorem \ref{thm:klp ck cg} actually holds for an arbitrary instance,
as stated formally below.
\begin{conjecture}
\label{conj:klp ck tightness}For any instance $\mathcal{I}=\left(a,v\right)$
(with items of arbitrary size) and any point $x\in P\left(a\right)$,
we have 
\[
\CG\left(a,v,x\right)\ge\frac{1-e^{-2}}{2}.
\]
\end{conjecture}

\subsection{\label{subsec:klp big small}Contention resolution schemes for instances
with only ``big'' or only ``small'' items}

In this section, we provide two contention resolution schemes for
the fractional knapsack polytope (introduced in Section \ref{sec:intro klp})
for two classes of instances: a monotone $\frac{1-e^{-2}}{2}$-balanced
CRS for class-$1$ instances (or, using the terminology from \cite{Bansal2012,Brubach2019},
instances with only ``big'' items) and a monotone $\frac{4}{9}$-balanced
CRS for instances where all items sizes are at most $\frac{1}{2}$
(i.e.,\,instances with only ``small'' items). These two schemes
constructively prove the lower bound on the correlation gap of class-$k$
knapsack instances we obtained in Theorem \ref{thm:klp ck cg}.

To simplify notation, throughout this section we use the shorthand
$a\left(R\right)\coloneqq\sum_{i\in R}a_{i}$ for any size vector
$a\in\left[0,1\right]^{n}$ and set $R\subseteq\left[n\right]$.
\begin{lem}
\label{lem:klp crs big}There exists a monotone $\frac{1-e^{-2}}{2}$-balanced
contention resolution scheme for the fractional knapsack polytope
for class-$1$ instances.
\end{lem}
\begin{proof}
We claim that Algorithm \ref{alg:klp crs big} is a monotone $\frac{1-e^{-2}}{2}$-balanced
CRS for the fractional knapsack polytope for class-$1$ instances.

\begin{algorithm2e}
\caption{\label{alg:klp crs big}Contention resolution scheme for the fractional knapsack polytope for class-$1$ instances.}
\KwIn{A size vector $a\in\left(\frac{1}{2},1\right]^{n}$, a point $x\in\left[0,1\right]^{n}$, a set $R\subseteq\supp\left(x\right)$.}
\KwOut{A point $y_{x}^{R}\in P_{I}\left(a\right)$ with $\supp\left(y_{x}^{R}\right)\subseteq R$.}
Sample each item $i\in R$ independently with probability $\frac{1-e^{-x_{i}}}{x_{i}}$ and let $\bar{R}$ be the set of sampled items\;
For each $i\in\bar{R}$, let $q_{i}\in\mathbb{Z}_{\ge1}$ be an independent realization of a $\Pois\left(x_{i}\right)$ random variable conditioned on it being at least $1$, and let $\left(y_{x}^{R}\right)_{i}\coloneqq\frac{q_{i}}{\sum_{j\in R}q_{j}}$\;
For each $i\notin\bar{R}$, let $q_{i}\coloneqq0$ and let $\left(y_{x}^{R}\right)_{i}\coloneqq0$\;
Return $y_{x}^{R}$\;
\end{algorithm2e}

Let $a\in\left(\frac{1}{2},1\right]^{n}$ be a size vector of $n$
class-$1$ items. Consider the graph $G=\left(V,E\right)$ with $V=\left\{ u\right\} \cup\left\{ t_{i}\right\} _{i=1}^{n}$
and $E=\left\{ \left(u,t_{i}\right)\mid i\in\left[n\right]\right\} $.
Let $\bar{x}_{\left(u,t_{i}\right)}\coloneqq x_{i}$ for every $i\in\left[n\right]$.
Since $x\in P\left(a\right)$ and $a_{i}>\frac{1}{2}$ for every $i\in\left[n\right]$,
we have 
\[
1\ge\sum_{i\in\left[n\right]}a_{i}x_{i}>\frac{1}{2}\sum_{e\in E}\bar{x}_{e},
\]
hence $\bar{x}\in2P\left(G\right)$, where $P\left(G\right)$ is the
fractional hypergraph matching polytope (see Section \ref{sec:hg}).
Even though $\bar{x}\in bP\left(G\right)$ with $b=2\notin\left[0,1\right]$,
note that applying Algorithm \ref{alg:klp crs big} to $a$, $x$,
and any set $R\subseteq\left[n\right]$ is equivalent to applying
Algorithm \ref{alg:hg gen crs} (or the original monotone CRS for
the general matching polytope presented in Section 4.1 of \cite{Bruggmann2019})
to $G$, $\bar{x}$, and the same set $R$, and, moreover, the analysis
in the proof of Theorem \ref{thm:hg gen} (or of Theorem 8 in \cite{Bruggmann2019})
goes through. Thus, correctness and monotonicity of Algorithm \ref{alg:klp crs big}
follow from correctness and monotonicity of Algorithm \ref{alg:hg gen crs}.
Since $\bar{x}\in2P\left(G\right)$, Theorem \ref{thm:hg gen} only
shows that our monotone CRS is $\frac{1-e^{-4}}{4}$-balanced. However,
we can exploit the structure of our graph $G$ to obtain an improved
lower bound on the balancedness (compared to the last series of inequalities
in the proof of Theorem \ref{thm:hg gen}):
\begin{align*}
\mathbb{E}\left[\left(y_{\bar{x}}\right)_{e}\right] & \ge\bar{x}_{e}\cdot\mathbb{E}\left[\frac{1}{1+\Pois\left(\sum_{f\in N\left(e\right)}\bar{x}_{f}\right)}\right]\\
 & =\bar{x}_{e}\cdot\mathbb{E}\left[\frac{1}{1+\Pois\left(\sum_{f\in E}\bar{x}_{f}\right)}\right]\\
 & \ge\bar{x}_{e}\cdot\mathbb{E}\left[\frac{1}{1+\Pois\left(2\right)}\right]\\
 & =\bar{x}_{e}\cdot\frac{1-e^{-2}}{2}.
\end{align*}
Thus, Algorithm \ref{alg:klp crs big} is a monotone $\frac{1-e^{-2}}{2}$-balanced
CRS.
\end{proof}
\begin{rem*}
Recall that in Section \ref{subsec:klp ck} we showed that for any
$a\in\left(\frac{1}{2},1\right]^{n}$ we have $\CG\left(P\left(a\right)\right)\ge\frac{1-e^{-2}}{2}$
(by Theorem \ref{thm:klp ck cg}) and there exist class-$1$ instances
for which the correlation gap is arbitrarily close to $\frac{1-e^{-2}}{2}$
(see Example \ref{exa:klp ck tight cg}). Thus, Algorithm \ref{alg:klp crs big}
has optimal balancedness.
\end{rem*}
\begin{lem}
\label{lem:klp crs small}There exists a monotone $\frac{4}{9}$-balanced
contention resolution scheme for the fractional knapsack polytope
for instances where all items sizes are at most $\frac{1}{2}$.
\end{lem}
\begin{proof}
We claim that Algorithm \ref{alg:klp crs small} is a monotone $\frac{4}{9}$-balanced
CRS for the fractional knapsack polytope.

\begin{algorithm2e}
\caption{\label{alg:klp crs small}Contention resolution scheme for the fractional knapsack polytope when all items are "small".}
\KwIn{A size vector $a\in\left[0,\frac{1}{2}\right]^{n}$, a point $x\in\left[0,1\right]^{n}$, a set $R\subseteq\supp\left(x\right)$.}
\KwOut{A point $y_{x}^{R}\in P_{I}\left(a\right)$ with $\supp\left(y_{x}^{R}\right)\subseteq R$.}
Set
\[y_{x}^{R}\coloneqq\begin{cases} \chi^{R}, & a\left(R\right)\le1,\\ \frac{2\chi^{R}}{3a\left(R\right)}, & a\left(R\right)>1; \end{cases}\]\DontPrintSemicolon\;\PrintSemicolon
Return $y_{x}^{R}$\;
\end{algorithm2e}

First, we prove that Algorithm \ref{alg:klp crs small} is indeed
a CRS by showing that it always returns $y_{x}^{R}\in P_{I}\left(a\right)$.
Clearly, $y_{x}^{R}\in\left[0,1\right]^{n}$. If $a\left(R\right)\le1$,
then $y_{x}^{R}=\chi^{R}\in P_{I}\left(a\right)$. Otherwise, we have
$\frac{\chi^{R}}{a\left(R\right)}\in P\left(a\right)$, since
\[
\sum_{j\in\left[n\right]}a_{j}\cdot\frac{\chi^{R}\left(j\right)}{a\left(R\right)}=\frac{a\left(R\right)}{a\left(R\right)}=1.
\]
Thus, by Lemma \ref{lem:tech klp dom} (since the greedy algorithm
gives a $\frac{3}{2}$-approximation by Lemma \ref{lem:ig klp small}
from Appendix \ref{subsec:ig klp}), there exists a set of $k\in\mathbb{Z}_{\ge1}$
integral points $z_{1},\dots,z_{k}\in P_{I}\left(a\right)$ and coefficients
$\lambda_{1},\dots,\lambda_{k}\in\mathbb{R}_{\ge0}$ such that $\sum_{i\in\left[k\right]}\lambda_{i}=1$
and $\frac{2\chi^{R}}{3a\left(R\right)}\le\sum_{i\in\left[k\right]}\lambda_{i}z_{i}$.
Since $P_{I}\left(a\right)$ is a down-closed polytope, we thus have
$y_{x}^{R}\in P_{I}\left(a\right)$, which proves correctness of Algorithm
\ref{alg:klp crs small}.

Next, we prove that our CRS is monotone by showing that $\left(y_{x}^{A}\right)_{i}\ge\left(y_{x}^{B}\right)_{i}$
for any point $x\in P\left(a\right)$ and item $i\in A\subseteq B\subseteq\supp\left(x\right)$
(note that the expectation is omitted, since the algorithm is deterministic).
Consider the following three cases.
\begin{enumerate}
\item If $a\left(B\right)\le1$, then $\left(y_{x}^{A}\right)_{i}=1=\left(y_{x}^{B}\right)_{i}$.
\item If $a\left(A\right)\le1<a\left(B\right)$, then $\left(y_{x}^{A}\right)_{i}=1\ge\frac{2}{3a\left(B\right)}=\left(y_{x}^{B}\right)_{i}$.
\item If $a\left(A\right)>1$, then $\left(y_{x}^{A}\right)_{i}=\frac{2}{3a\left(A\right)}\ge\frac{2}{3a\left(B\right)}=\left(y_{x}^{B}\right)_{i}$.
\end{enumerate}
Thus, we have $\left(y_{x}^{A}\right)_{i}\ge\left(y_{x}^{B}\right)_{i}$,
so Algorithm \ref{alg:klp crs small} is a monotone CRS.

Finally, we prove that our CRS is $\frac{4}{9}$-balanced. Let $R\subseteq\left[n\right]$
be a random set where each item $i\in\left[n\right]$ is included
independently with probability $x_{i}$. To simplify notation, we
use $\mathbb{E}_{\le1}$ and $\mathbb{E}_{>1}$ to denote expectations
with additional conditioning on $a\left(R\right)\le1$ and $a\left(R\right)>1$,
respectively. We also denote
\[
p_{j}\coloneqq\Pr\left[a\left(R\right)>1\mid j\in R\right].
\]
Then for any $j\in\left[n\right]$, we have
\[
\mathbb{E}\left[\left(y_{x}^{R}\right)_{j}\mid j\in R\right]=\left(1-p_{j}\right)\cdot\mathbb{E}_{\le1}\left[\left(y_{x}^{R}\right)_{j}\mid j\in R\right]+p_{j}\cdot\mathbb{E}_{>1}\left[\left(y_{x}^{R}\right)_{j}\mid j\in R\right].
\]
Consider the following two cases.
\begin{enumerate}
\item Suppose that $a\left(R\right)\le1$. Then $\left(y_{x}^{R}\right)_{j}=1$
for any $j\in R$, hence 
\[
\mathbb{E}_{\le1}\left[\left(y_{x}^{R}\right)_{j}\mid j\in R\right]=1.
\]
\item Suppose that $a\left(R\right)>1$. Then for any $j\in R$, we have
\begin{align*}
\mathbb{E}_{>1}\left[\left(y_{x}^{R}\right)_{j}\mid j\in R\right] & =\frac{2}{3}\cdot\mathbb{E}_{>1}\left[\frac{1}{a_{j}+a\left(R\setminus\left\{ j\right\} \right)}\mid j\in R\right]\\
 & \ge\frac{2}{3}\cdot\frac{1}{a_{j}+\mathbb{E}_{>1}\left[a\left(R\setminus\left\{ j\right\} \right)\mid j\in R\right]}\\
 & \ge\frac{2}{3}\cdot\frac{p_{j}}{a_{j}p_{j}+\sum_{i\in\left[n\right]\setminus\left\{ j\right\} }a_{i}x_{i}}\\
 & \ge\frac{2}{3}\cdot\frac{p_{j}}{a_{j}\left(1-x_{j}\right)+1}\\
 & \ge\frac{4p_{j}}{9}.
\end{align*}
The first inequality holds by Jensen's inequality for conditional
expectations (with $X\coloneqq\sum_{i\in R\setminus\left\{ j\right\} }a_{i}$
as the random variable and $\varphi\left(X\right)\coloneqq\frac{1}{a_{j}+X}$
as the convex function). The second inequality holds, since
\[
\sum_{i\in\left[n\right]\setminus\left\{ j\right\} }a_{i}x_{i}=\mathbb{E}\left[a\left(R\setminus\left\{ j\right\} \right)\right]\ge p_{j}\mathbb{E}_{>1}\left[a\left(R\setminus\left\{ j\right\} \right)\mid j\in R\right].
\]
The third inequality holds, since $p_{j}\le1$ and $\sum_{i\in\left[n\right]\setminus\left\{ j\right\} }a_{i}x_{i}\le1-a_{j}x_{j}$.
The last inequality holds, since $a_{j}\left(1-x_{j}\right)\le\frac{1}{2}$.
\end{enumerate}
Combining the results, we get:
\begin{align*}
\mathbb{E}\left[\left(y_{x}^{R}\right)_{j}\mid j\in R\right] & \ge\left(1-p_{j}\right)\cdot1+p_{j}\cdot\frac{4p_{j}}{9}\\
 & =\frac{4}{9}p_{j}^{2}-p_{j}+1\\
 & \ge\min_{p\in\left[0,1\right]}\left(\frac{4}{9}p^{2}-p+1\right)\\
 & =\frac{4}{9},
\end{align*}
where the last equality holds, since the constrained minimum is attained
at $p=1$. Thus, Algorithm \ref{alg:klp crs gen} is a monotone $\frac{4}{9}$-balanced
CRS.
\end{proof}
\begin{rem*}
Note that the bound $\mathbb{E}\left[\left(y_{x}^{R}\right)_{j}\mid j\in R\right]\ge\frac{4}{9}$
in the proof of Lemma \ref{lem:klp crs small} is tight: consider
an instance with $a=\frac{1}{2+\varepsilon}\cdot1_{3}$ and $x=\left(\varepsilon,1,1\right)^{\top}$
where $\varepsilon>0$ is arbitrarily small; then
\[
\frac{\mathbb{E}\left[\left(y_{x}^{R}\right)_{1}\right]}{x_{1}}=\frac{2}{3\cdot\frac{3}{2+\varepsilon}}=\xrightarrow{\varepsilon\to+0}\frac{4}{9}.
\]
\end{rem*}
\begin{cor*}
Since $\frac{4}{9}>\frac{1-e^{-2}}{2}$, Lemmas \ref{lem:klp crs big}
and \ref{lem:klp crs small} provide a constructive proof of the special
case of Conjecture \ref{conj:klp ck tightness} where the knapsack
instance consists either of only ``big'' or of only ``small''
items, which includes Theorem \ref{thm:klp ck cg}.
\end{cor*}

\subsection{\label{subsec:klp gen}Contention resolution schemes for general
instances}

In this section, we develop a monotone $0.279$-balanced CRS for the
fractional knapsack polytope (with arbitrary item sizes). The main
idea is to probabilistically combine two contention resolution schemes:
one that works only for class-$1$ items and has balancedness higher
than the required $0.279$ (i.e.,\,Algorithm \ref{alg:klp crs big}),
and the other that works for all items, but has lower balancedness
(i.e.,\,Algorithm \ref{alg:klp crs gen}, which is a slight modification
of Algorithm \ref{alg:klp crs small}). The improvement comes from
the separate analysis of the balancedness for ``big'' and ``small''
items (i.e.,\,items of size $a_{j}>\frac{1}{2}$ and of size $a_{j}\le\frac{1}{2}$,
respectively). To simplify notation, throughout this section we use
the shorthand $a\left(R\right)\coloneqq\sum_{i\in R}a_{i}$ for any
size vector $a\in\left[0,1\right]^{n}$ and set $R\subseteq\left[n\right]$.

We begin with a simple monotone $\frac{1}{4}$-balanced CRS for general-case
instances.
\begin{lem}
\label{lem:klp gen}There exists a monotone $\frac{1}{4}$-balanced
contention resolution scheme for the fractional knapsack polytope.
\end{lem}
\begin{proof}
We claim that Algorithm \ref{alg:klp crs gen} is a monotone $\frac{1}{4}$-balanced
CRS for the fractional knapsack polytope.

\begin{algorithm2e}
\caption{\label{alg:klp crs gen}Contention resolution scheme for the fractional knapsack polytope.}
\KwIn{A size vector $a\in\left[0,1\right]^{n}$, a point $x\in\left[0,1\right]^{n}$, a set $R\subseteq\supp\left(x\right)$.}
\KwOut{A point $y_{x}^{R}\in P_{I}\left(a\right)$ with $\supp\left(y_{x}^{R}\right)\subseteq R$.}
Set \[y_{x}^{R}\coloneqq\begin{cases} \chi^{R}, & a\left(R\right)\le1,\\ \frac{\chi^{R}}{2a\left(R\right)}, & a\left(R\right)>1; \end{cases}\]\DontPrintSemicolon\;\PrintSemicolon
Return $y_{x}^{R}$\;
\end{algorithm2e}

Note that Algorithms \ref{alg:klp crs small} and \ref{alg:klp crs gen}
only differ in the scaling factor that is used when $a\left(R\right)>1$,
so the proof of this lemma is almost exactly the same as that of Lemma
\ref{lem:klp crs big}. By replacing the factor $\frac{3}{2}$ by
$2$ in the proof of correctness and monotonicity of Algorithm \ref{alg:klp crs small}
from Lemma \ref{lem:klp crs big}, we get a proof of correctness and
monotonicity of Algorithm \ref{alg:klp crs gen}. Similarly, the following
changes to the analysis of balancedness of Algorithm \ref{alg:klp crs small}
from Lemma \ref{lem:klp crs big} prove $\frac{1}{4}$-balancedness
of Algorithm \ref{alg:klp crs gen}:
\[
\mathbb{E}_{>1}\left[\left(y_{x}^{R}\right)_{j}\mid j\in R\right]\ge\frac{1}{2}\cdot\frac{p_{j}}{a_{j}\left(1-x_{j}\right)+1}\ge\frac{p_{j}}{4},
\]
\begin{align*}
\mathbb{E}\left[\left(y_{x}^{R}\right)_{j}\mid j\in R\right] & \ge\left(1-p_{j}\right)\cdot1+p_{j}\cdot\frac{p_{j}}{4}\\
 & \ge\min_{p\in\left[0,1\right]}\left(\frac{p^{2}}{4}-p+1\right)\\
 & =\frac{1}{4}.
\end{align*}
Thus, Algorithm \ref{alg:klp crs gen} is a monotone $\frac{1}{4}$-balanced
CRS.
\end{proof}
\begin{rem*}
Note that the bound $\mathbb{E}\left[\left(y_{x}^{R}\right)_{j}\mid j\in R\right]\ge\frac{1}{4}$
in the proof of Lemma \ref{lem:klp gen} is tight: consider an instance
with $a=\frac{1}{1+\varepsilon}\cdot1_{2}$ and $x=\left(\varepsilon,1\right)^{\top}$
where $\varepsilon>0$ is arbitrarily small; then
\[
\frac{\mathbb{E}\left[\left(y_{x}^{R}\right)_{1}\right]}{x_{1}}=\frac{1}{2\cdot\frac{2}{1+\varepsilon}}\xrightarrow{\varepsilon\to+0}\frac{1}{4}.
\]
\end{rem*}
The following lemma shows that, roughly speaking, Algorithm \ref{alg:klp crs gen}
is $\frac{1}{3}$-balanced with respect to ``small'' items in contrast
to it being $\frac{1}{4}$-balanced with respect to ``big'' items
as shown above in Lemma \ref{lem:klp gen}.
\begin{lem}
\label{lem:klp gen crs for small}In the setting of Lemma \ref{lem:klp gen},
for any item $j\in\left[n\right]$ with $a_{j}\le\frac{1}{2}$, the
output of Algorithm \ref{alg:klp crs gen} satisfies
\[
\mathbb{E}\left[\left(y_{x}^{R}\right)_{j}\mid j\in R\right]\ge\frac{1}{3}.
\]
\end{lem}
\begin{proof}
Here we use the same notation as in the proof of Lemma \ref{lem:klp gen}.
Since $a_{j}\left(1-x_{j}\right)\le\frac{1}{2}$, we have
\[
\mathbb{E}_{>1}\left[\left(y_{x}^{R}\right)_{j}\mid j\in R\right]\ge\frac{1}{2}\cdot\frac{p_{j}}{a_{j}\left(1-x_{j}\right)+1}\ge\frac{p_{j}}{3},
\]
so
\begin{align*}
\mathbb{E}\left[\left(y_{x}^{R}\right)_{j}\mid j\in R\right] & \ge\left(1-p_{j}\right)\cdot1+p_{j}\cdot\frac{p_{j}}{3}\\
 & \ge\min_{p\in\left[0,1\right]}\left(\frac{p^{2}}{3}-p+1\right)\\
 & \ge\frac{1}{3}.
\end{align*}
\end{proof}
\begin{rem*}
The bound $\mathbb{E}\left[\left(y_{x}^{R}\right)_{j}\mid j\in R\right]\ge\frac{1}{3}$
for items $j\in\left[n\right]$ with $a_{j}\le\frac{1}{2}$ is, in
fact, tight: consider an instance with $a=\left(\frac{1-\varepsilon}{2},\frac{1-\varepsilon}{2},\frac{1}{2}\right)^{\top}$
and $x=\left(1,1,2\varepsilon\right)^{\top}$ where $\varepsilon>0$
is arbitrarily small; then 
\[
\frac{\mathbb{E}\left[\left(y_{x}^{R}\right)_{3}\right]}{x_{3}}=\frac{1}{2\left(2\cdot\frac{1-\varepsilon}{2}+\frac{1}{2}\right)}=\frac{1}{3-2\varepsilon}\xrightarrow{\varepsilon\to+0}\frac{1}{3}.
\]
\end{rem*}
Now we probabilistically combine Algorithms \ref{alg:klp crs big}
and \ref{alg:klp crs gen} to obtain a monotone $0.279$-balanced
CRS for the fractional knapsack polytope (for general instances).
\begin{lem}
There exists a monotone $0.279$-balanced contention resolution scheme
for the fractional knapsack polytope.
\end{lem}
\begin{proof}
We claim that Algorithm \ref{alg:klp crs gen imp} where we set $q\coloneqq\frac{6\left(e^{2}-1\right)}{7e^{2}-6}\approx0.838$
is a monotone $0.279$-balanced CRS for the fractional knapsack polytope.

\begin{algorithm2e}
\caption{\label{alg:klp crs gen imp}Improved contention resolution scheme for the fractional knapsack polytope.}
\KwIn{A size vector $a\in\left[0,1\right]^{n}$, a point $x\in\left[0,1\right]^{n}$, a set $R\subseteq\supp\left(x\right)$.}
\KwOut{A point $y_{x}^{R}\in P_{I}\left(a\right)$ with $\supp\left(y_{x}^{R}\right)\subseteq R$.}
With probability $q\coloneqq\frac{6\left(e^{2}-1\right)}{7e^{2}-6}$, let $y_{x}^{R}$ be the output of Algorithm \ref{alg:klp crs gen} applied to $a$, $x$, and $R$\;
Otherwise (with probability $1-q$), let $\bar{R}\subseteq R$ be the set of all class-$1$ items in $R$, and let $y_{x}^{R}$ be the output of Algorithm \ref{alg:klp crs big} applied to $a$, $x$, and $\bar{R}$\;
Return $y_{x}^{R}$\;
\end{algorithm2e}

Correctness and monotonicity of Algorithm \ref{alg:klp crs gen imp}
follow from correctness and monotonicity of Algorithms \ref{alg:klp crs big}
and \ref{alg:klp crs gen}. Let $R\subseteq\left[n\right]$ be a random
subset where each item $i\in\left[n\right]$ is included independently
with probability $x_{i}$, and let $y_{x}^{R}$ be the output of Algorithm
\ref{alg:klp crs gen imp}. Then, by Theorem \ref{thm:hg gen} and
Lemma \ref{lem:klp gen}, for any class-$1$ item $j\in\left[n\right]$,
we have
\begin{align*}
\mathbb{E}\left[\left(y_{x}^{R}\right)_{j}\mid j\in R\right] & \ge q\cdot\frac{1}{4}+\left(1-q\right)\cdot\frac{1-e^{-2}}{2}\\
 & =\frac{1-e^{-2}}{2}-\frac{1-2e^{-2}}{4}q,
\end{align*}
and by Lemma \ref{lem:klp gen crs for small}, for any item $j\in\left[n\right]$
such that $a_{j}\le\frac{1}{2}$, we have
\[
\mathbb{E}\left[\left(y_{x}^{R}\right)_{j}\mid j\in R\right]\ge q\cdot\frac{1}{3}.
\]
Thus, for any $j\in\left[n\right]$, we have
\[
\mathbb{E}\left[\left(y_{x}^{R}\right)_{j}\mid j\in R\right]\ge\min\left\{ \frac{1-e^{-2}}{2}-\frac{1-2e^{-2}}{4}q,\frac{q}{3}\right\} .
\]
By setting $q\coloneqq\frac{6\left(e^{2}-1\right)}{7e^{2}-6}\approx0.838$,
we make the two values in the minimum equal and get 
\[
\mathbb{E}\left[\left(y_{x}^{R}\right)_{j}\mid j\in R\right]\ge\frac{2\left(e^{2}-1\right)}{7e^{2}-6}\ge0.279,
\]
which proves the theorem.
\end{proof}

\newpage{}

\section{\label{sec:kcs}Contention resolution scheme for the natural LP relaxation
of $k$-CS-PIP with refined analysis}

In this section, we give a $\frac{1}{4k+o\left(k\right)}$-balanced
CRS for the natural LP relaxation of $k$-CS-PIP (introduced in Section
\ref{sec:intro kcs}) by slightly modifying the $\left(2k+o\left(k\right)\right)$-approximation
algorithm from \cite{Brubach2019} (namely, Algorithm 2), which is
based on the strengthened LP relaxation. In addition, we slightly
refine the analysis of the algorithm and improve the $o\left(k\right)$
term in the approximation ratio. Combining this scheme with a sampling
step (where each item $j\in\left[n\right]$ is independently sampled
with probability $x_{j}$) gives a $\left(4k+o\left(k\right)\right)$-approximation
algorithm for $k$-CS-PIP based on the natural LP relaxation, thus
asymptotically improving the $8k$-approximation algorithm given in
\cite{Bansal2012}. We use the same notation as \cite{Brubach2019}.
\begin{defn}[Item classes and blocking events]
\label{def:kcs item classes}Let $\ell\ge2$ be a given real parameter.
Let 
\begin{align*}
\mathrm{big}\left(i\right) & \coloneqq\left\{ j\in\left[n\right]\mid a_{ij}>\frac{1}{2}\right\} \ \forall i\in\left[m\right],\\
\mathrm{med}\left(i\right) & \coloneqq\left\{ j\in\left[n\right]\mid\frac{1}{\ell}<a_{ij}\le\frac{1}{2}\right\} \ \forall i\in\left[m\right],\\
\mathrm{tiny}\left(i\right) & \coloneqq\left\{ j\in\left[n\right]\mid0<a_{ij}\le\frac{1}{\ell}\right\} \ \forall i\in\left[m\right]
\end{align*}
denote the sets of big, medium, and tiny items with respect to each
constraint. For a randomly sampled set $R\subseteq\left[n\right]$
and an item $j\in R$, we distinguish between three kinds of blocking
events for $j$ with respect to $R$.
\begin{itemize}
\item Big blocking event $\mathrm{BB}\left(j,R\right)$: there exists a
constraint $i\in\left[m\right]$ such that $a_{ij}>0$ and there exists
an item $j'\in R\cap\mathrm{big}\left(i\right)\setminus\left\{ j\right\} $.
\item Medium blocking event $\mathrm{MB}\left(j,R\right)$: there exists
a constraint $i\in\left[m\right]$ such that $j\in\mathrm{med}\left(i\right)$
and $\left|R\cap\mathrm{med}\left(i\right)\right|\ge3$.
\item Tiny blocking event $\mathrm{TB}\left(j,R\right)$: there exists a
constraint $i\in\left[m\right]$ such that $j\in\mathrm{tiny}\left(i\right)$
and
\[
\left|R\cap\mathrm{med}\left(i\right)\right|\ge2\quad\text{or}\quad\sum_{j'\in R\cap\left(\mathrm{med}\left(i\right)\cup\mathrm{tiny}\left(i\right)\right)}a_{ij'}>1.
\]
\end{itemize}
Each of these events can potentially prevent the set $R$ from being
rounded up without violating some constraints.
\end{defn}
\begin{algorithm2e}
\caption{\label{alg:kcs crs}Contention resolution scheme for the natural LP relaxation of $k$-CS-PIP.}
\KwIn{A $k$-column sparse constraint matrix $a\in\left[0,1\right]^{m\times n}$, a point $x\in\left[0,1\right]^{n}$ such that $\sum_{j\in\left[n\right]}a_{ij}x_{j}\le1$ for every $i\in\left[m\right]$, a set $R\subseteq\supp\left(x\right)$.}
\KwOut{A set $R_{F}\subseteq R$ such that $\sum_{j\in R_{F}}a_{ij}\le1$ for every $i\in\left[m\right]$.}
{\bf Sampling:} Sample each item $j$ independently with probability $\frac{\alpha}{k}$ where $\alpha\coloneqq k^{0.4}$ and let $R_{0}$ be the set of sampled items\;
{\bf Discarding low-probability events:} Remove an item $j$ from $R_{0}$ if either a medium or tiny blocking event occurs for $j$ with respect to $R_{0}$ and let $R_{1}$ be the set of items not removed\;
{\bf Creating a directed graph:} For every item $j\in R_{1}$, create a vertex in a graph $G$ and add a directed edge to every item $j'\neq j$ for which there exists a constraint $i$ such that $a_{ij}>0$ and $a_{ij}>\frac{1}{2}$\;
{\bf Removing anomalous vertices:} For every vertex $v\in G$, if the out-degree of $v$ is greater than $d\coloneqq 2\alpha+\sqrt{\alpha\log\alpha}$, call $v$ anomalous. Remove all anomalous vertices from $G$ to obtain $G'$ and let $R_{2}$ be the items corresponding to the remaining vertices in $G'$\;
{\bf Coloring G':} Assign a coloring $\chi$ to the vertices of $G'$ using $2d+1$ colors such that endpoints of every edge $e$ (ignoring the direction) receive different colors\;
{\bf Choosing an independent set:} Choose a number $c\in\left[2d+1\right]$ uniformly at random and let $R_{F}$ be all vertices $v\in G'$ such that $\chi\left(v\right)=c$\;
Return $R_{F}$\;
\end{algorithm2e}
\begin{thm}
\label{thm:kcs approx crs}Algorithm \ref{alg:kcs crs} is a $\frac{1}{4k+o\left(k\right)}$-balanced
CRS for the natural LP relaxation of $k$-CS-PIP.
\end{thm}
\begin{proof}
The proof of this theorem closely resembles that of Theorem 1 in \cite{Brubach2019}.

Clearly, Algorithm \ref{alg:kcs crs} is a CRS, since the returned
set $R_{F}$ is feasible by construction. Let $R\subseteq\left[n\right]$
be a random set where each item $j\in\left[n\right]$ is included
independently with probability $x_{j}$. Then for every $j\in\left[n\right]$
we have
\begin{align*}
\mathbb{E}\left[j\in R_{F}\right] & =\Pr\left[j\in R\right]\cdot\Pr\left[j\in R_{0}\mid j\in R\right]\cdot\Pr\left[j\in R_{1}\mid j\in R_{0}\right]\cdot\Pr\left[j\in R_{2}\mid j\in R_{1}\right]\cdot\Pr\left[j\in R_{F}\mid j\in R_{2}\right]\\
 & \ge x_{j}\cdot\frac{\alpha}{k}\cdot\left(1-o\left(1\right)\right)\cdot\left(1-o\left(1\right)\right)\cdot\frac{1}{4\alpha+2\sqrt{\alpha\log\alpha}+1}\\
 & =\frac{x_{j}}{4k+o\left(k\right)}.
\end{align*}
Here the inequality is a combination of the following four statements:
\begin{enumerate}
\item from the sampling step, we have 
\[
\Pr\left[j\in R_{0}\right]=\frac{\alpha}{k};
\]
\item by Lemma \ref{lem:kcs med tiny}, which is identical to Lemma 3 from
\cite{Brubach2019}, we have 
\[
\Pr\left[j\in R_{1}\mid j\in R_{0}\right]\ge1-o\left(1\right);
\]
\item by Lemma \ref{lem:kcs anomalous}, which is very similar to Lemma
2 from \cite{Brubach2019}, we have 
\[
\Pr\left[j\in R_{2}\mid j\in R_{1}\right]\ge1-o\left(1\right);
\]
\item by Lemma \ref{lem:kcs coloring}, which is identical to Lemma 1 from
\cite{Brubach2019}, we have
\[
\Pr\left[j\in R_{F}\mid j\in R_{2}\right]=\frac{1}{2d+1}=\frac{1}{4\alpha+2\sqrt{\alpha\log\alpha}+1}.
\]
\end{enumerate}
Therefore, Algorithm \ref{alg:kcs crs} is a $\frac{1}{4k+o\left(k\right)}$-balanced
CRS. Moreover, the analysis in Lemmas \ref{lem:kcs med tiny} and
\ref{lem:kcs anomalous} also goes through if we set $\alpha\coloneqq\alpha\left(k\right)$,
$\ell\coloneqq\beta\left(k\right)$, and $d\coloneqq2\alpha+\sqrt{\alpha}\gamma\left(k\right)$
where $\alpha\left(k\right)$, $\beta\left(k\right)$ and $\gamma\left(k\right)$
are arbitrarily slowly growing functions of $k$ such that $\alpha\left(k\right)<k^{0.5}$
and $k$ is large enough.
\end{proof}
\begin{cor}
Algorithm \ref{alg:kcs crs} applied to a random set $R\subseteq\left[n\right]$
where each item $j\in\left[n\right]$ is included with probability
$x_{j}$ independently gives a $\left(4k+o\left(k\right)\right)$-approximation
for the natural LP relaxation of $k$-CS-PIP. This is an asymptotic
improvement over the $8k$-approximation algorithm given in \cite{Bansal2012}.
However, it does not settle the question about the integrality gap
of the natural LP relaxation, since the best currently known lower
bound is $2k-1$, attained by an example from \cite{Bansal2012} (which
we provide in Appendix \ref{subsec:ig kcs} for the sake of convenience).
\end{cor}
\begin{lem}
\label{lem:kcs med tiny}For every item $j\in\left[n\right]$, we
have
\[
\Pr\left[\mathrm{MB}\left(j,R_{0}\right)\wedge\mathrm{TB}\left(j,R_{0}\right)\mid j\in R_{0}\right]\le o\left(1\right),
\]
where the probability is computed with respect to random sampling
of $R\subseteq\supp\left(x\right)$ and $R_{0}\subseteq R$.
\end{lem}
\begin{proof}
The proof of this lemma closely resembles that of Lemma 3 from \cite{Brubach2019}.

First, we introduce the necessary notation and make several useful
observations. For every item $j'\in\left[n\right]$, let $X_{j'}$
denote the random variable indicating that $j'\in R_{0}$. To simplify
notation, for any constraint $i\in\left[m\right]$, let $G_{i}\coloneqq\mathrm{med}\left(i\right)\cap R_{0}\setminus\left\{ j\right\} $
and $H_{i}\coloneqq\mathrm{tiny}\left(i\right)\cap R_{0}\setminus\left\{ j\right\} $.
For medium items, the constraint $i$ implies that 
\[
\mathbb{E}\left[\sum_{g\in G_{i}}X_{g}\right]\le\frac{\alpha}{k}\sum_{g\in\mathrm{med}\left(i\right)}x_{g}\le\frac{\alpha\ell}{k}\sum_{g\in\mathrm{med}\left(i\right)}a_{ig}x_{g}\le\frac{\alpha\ell}{k}.
\]
For every $h\in H_{i}$, define a random variable $Y_{h}\coloneqq\ell a_{ih}X_{h}$
and observe that $Y_{h}\in\left[0,1\right]$ and
\[
\mathbb{E}\left[\sum_{h\in H_{i}}Y_{h}\right]\le\frac{\alpha\ell}{k}\sum_{h\in\mathrm{tiny}\left(i\right)}a_{ih}x_{h}\le\frac{\alpha\ell}{k}.
\]

Now, consider the medium blocking event $\mathrm{MB}\left(j,R_{0}\right)$
and let $i\in\left[m\right]$ be a constraint causing this event.
The event $\mathrm{MB}\left(j,R_{0}\right)$ might prevent the item
$j$ from being rounded up if $\sum_{g\in G_{i}}X_{g}\ge2$. Under
the condition that $\frac{\alpha\ell}{k}\ge2$, by Lemma \ref{lem:tech kcs pow},
we have
\[
\Pr\left[\sum_{g\in G_{i}}X_{g}\ge2\mid X_{j}=1\right]=\Pr\left[\sum_{g\in G_{i}}X_{g}\ge2\right]\le\left(\frac{\alpha\ell e}{2k}\right)^{2}=\frac{e^{2}}{4}\cdot\frac{\alpha^{2}\ell^{2}}{k}.
\]
Thus, by a union bound over at most $k$ constraints that the item
$j$ participates in, the total probability that the event $\mathrm{MB}\left(j,R_{0}\right)$
might prevent the item $j$ from being rounded up is at most $\frac{e^{2}}{4}\cdot\frac{\alpha^{2}\ell^{2}}{k}$.

Finally, consider the tiny blocking event $\mathrm{TB}\left(j,R_{0}\right)$
and let $i\in\left[m\right]$ be a constraint causing this event.
We consider the following three cases that are sufficient to potentially
cause $\mathrm{TB}\left(j,R_{0}\right)$ to prevent the item $j$
from being rounded up.
\begin{enumerate}
\item Suppose that there exist at least two items $g_{1},g_{2}\in G_{i}$.
In this case, the item $j$ might be prevented from being rounded
up if $\sum_{g\in G_{i}}X_{g}\ge2$, which is the same as for the
medium blocking event. Thus, the probability that $j$ might be prevented
from being rounded up is at most $\frac{e^{2}}{4}\cdot\frac{\alpha^{2}\ell^{2}}{k}$.
\item Suppose that there exists an item $g\in G_{i}$ such that
\[
a_{ig}+\sum_{h\in H_{i}\cup\left\{ j\right\} }a_{ih}>1.
\]
Define $\delta_{0}\coloneqq\frac{k\left(\ell-2\right)}{2\alpha\ell}-1$
(which satisfies $\left(1+\delta_{0}\right)\frac{\alpha\ell}{k}=\frac{\ell}{2}-1$)
and $\delta\coloneqq\frac{k\left(\ell-2\right)}{4\alpha\ell}$. In
this case, the probability that the item $j$ is prevented from being
rounded up is upper bounded by 
\begin{align*}
\Pr\left[\sum_{h\in H_{i}}a_{ih}X_{h}>1-a_{ig}-a_{ij}\right] & \le\Pr\left[\sum_{h\in H_{i}}a_{ih}X_{h}>\frac{1}{2}-\frac{1}{\ell}\right]\\
 & =\Pr\left[\sum_{h\in H_{i}}Y_{h}>\frac{\ell}{2}-1\right]\\
 & \le\Pr\left[\sum_{h\in H_{i}}Y_{h}>\left(1+\delta\right)\frac{\alpha\ell}{k}\right]\\
 & \le\exp\left(-\frac{\delta^{2}}{2+\delta}\cdot\frac{\alpha\ell}{k}\right)\\
 & =\exp\left(-\frac{k\left(\ell-2\right)^{2}}{32\alpha\ell+4k\left(\ell-2\right)}\right)\\
 & \le\exp\left(-\frac{\left(\ell-2\right)^{2}}{36\ell-8}\right)\\
 & \le\exp\left(-\frac{\ell}{300}\right).
\end{align*}
The first inequality holds, since $j\in\mathrm{tiny}\left(i\right)$
and $g\in\mathrm{med}\left(i\right)$. The second inequality holds
if $\delta\le\delta_{0}$, which holds under the condition that $\ell\ge3$
and $\alpha\le\frac{k}{12}$ (note that this is a sufficient, but
not a necessary condition). The third inequality holds by Lemma \ref{lem:tech kcs mul}.
The fourth inequality holds, since we assume that $\alpha\le k$.
The last inequality holds under the condition that $\ell\ge3$ (the
original paper contains a computational error; by using the inequalities
$\frac{1}{2}\ge\frac{\ell/2-1}{\ell}\ge\frac{1}{6}$, one would get
$\frac{\delta^{2}}{2+\delta}\cdot\frac{\alpha\ell}{k}\ge\frac{\ell}{324}$).
\item Suppose that
\[
\sum_{h\in H_{i}\cup\left\{ j\right\} }a_{ih}>1.
\]
Define $\delta_{0}\coloneqq\frac{k\left(\ell-1\right)}{\alpha\ell}-1$
(which satisfies $\left(1+\delta_{0}\right)\frac{\alpha\ell}{k}=\ell-1$)
and $\delta\coloneqq\frac{k\left(\ell-1\right)}{2\alpha\ell}$. In
this case, the probability that the item $j$ is prevented from being
rounded up is upper bounded by
\begin{align*}
\Pr\left[\sum_{h\in H_{i}}a_{ih}X_{h}>1-a_{ij}\right] & \le\Pr\left[\sum_{h\in H_{i}}a_{ih}X_{h}>1-\frac{1}{\ell}\right]\\
 & =\Pr\left[\sum_{h\in H_{i}}Y_{h}>\ell-1\right]\\
 & \le\Pr\left[\sum_{h\in H_{i}}Y_{h}>\left(1+\delta\right)\frac{\alpha\ell}{k}\right]\\
 & \le\exp\left(-\frac{\delta^{2}}{2+\delta}\cdot\frac{\alpha\ell}{k}\right)\\
 & =\exp\left(-\frac{k\left(\ell-1\right)^{2}}{8\alpha\ell+2k\left(\ell-1\right)}\right)\\
 & \le\exp\left(-\frac{\left(\ell-1\right)^{2}}{10\ell-2}\right)\\
 & \le\exp\left(-\frac{\ell}{36}\right).
\end{align*}
The first inequality holds, since $j\in\mathrm{tiny}\left(i\right)$.
The second inequality holds if $\delta\le\delta_{0}$, which holds
under the condition that $\ell\ge2$ and $\alpha\le\frac{k}{4}$ (note
that this is a sufficient, but not a necessary condition). The third
inequality holds by Lemma \ref{lem:tech kcs mul}. The fourth inequality
holds, since we assume that $\alpha\le k$. The last inequality holds
under the condition that $\ell\ge2$.
\end{enumerate}
By summarizing the results, we obtain:
\[
\Pr\left[\mathrm{MB}\left(j,R_{0}\right)\wedge\mathrm{TB}\left(j,R_{0}\right)\mid j\in R_{0}\right]\le\min\left\{ \frac{e^{2}}{4}\cdot\frac{\alpha^{2}\ell^{2}}{k},\exp\left(-\frac{\ell}{300}\right),\exp\left(-\frac{\ell}{36}\right)\right\} .
\]
This upper bound is $o\left(1\right)$ if we set $\alpha\coloneqq k^{0.4}$
and $\ell\coloneqq\log k$.
\end{proof}
\begin{lem}
\label{lem:kcs anomalous}For every item $j\in\left[n\right]$, we
have 
\[
\Pr\left[j\in R_{2}\mid j\in R_{1}\right]=1-o\left(1\right),
\]
where the probability is computed with respect to random sampling
of $R\subseteq\supp\left(x\right)$ and $R_{0}\subseteq R$.
\end{lem}
\begin{proof}
The proof of this lemma closely resembles that of Lemma 2 from \cite{Brubach2019}.

Consider an item $j\in R_{1}$. Let $\delta_{j}$ be the out-degree
of the vertex corresponding to $j$ in the directed graph $G$. Recall
from the construction of the graph $G$ that we have an arc $\left(j,j'\right)\in E$
if and only if there is a constraint $i\in\left[m\right]$ such that
$a_{ij'}>\frac{1}{2}$ and $a_{ij}>0$. Let $\mathcal{S}\coloneqq\left\{ j'\neq j\mid\left(j,j'\right)\in E\right\} $,
and for every $j'\in\mathcal{S}$ let $X_{j'}\coloneqq\bbone\left[j'\in R_{1}\right]$.
Then by Lemma \ref{lem:kcs med tiny}, for any $j'\in\mathcal{S}$
we have
\begin{align*}
\mathbb{E}\left[X_{j'}\right] & =\Pr\left[j'\in R_{1}\right]\\
 & =\Pr\left[j'\in R\right]\cdot\Pr\left[j'\in R_{0}\mid j'\in R\right]\cdot\Pr\left[j'\in R_{1}\mid j'\in R_{0}\right]\\
 & =x_{j}\cdot\frac{\alpha}{k}\cdot\left(1-o\left(1\right)\right).
\end{align*}
Therefore,
\[
\mathbb{E}\left[\delta_{j}\right]=\mathbb{E}\left[\sum_{j'\in\mathcal{S}}X_{j'}\right]=\frac{\alpha}{k}\sum_{j'\in\mathcal{S}}x_{j'}\left(1-o\left(1\right)\right).
\]
Since the constraint matrix is $k$-column sparse, items $j'\in\mathcal{S}$
cause big blocking events with respect to at most $k$ constraints.
Since
\[
\frac{1}{2}\cdot\sum_{j'\in\mathrm{big}\left(i\right)}x_{j'}<\sum_{j'\in\mathrm{big}\left(i\right)}a_{ij'}x_{j'}\le\sum_{j'\in\left[n\right]}a_{ij'}x_{j'}\le1\quad\forall i\in\left[m\right],
\]
we have $\sum_{j'\in\mathrm{big}\left(i\right)}x_{j'}\le2$ for every
$i\in\left[m\right]$, so $\mathbb{E}\left[\delta_{j}\right]\le\left(2-o\left(1\right)\right)\alpha\le2\alpha$.
Therefore,
\begin{align*}
\Pr\left[\delta_{j}>d\mid j\in R_{1}\right] & =\Pr\left[\delta_{j}>2\alpha+\sqrt{\alpha\log\alpha}\right]\\
 & \le\Pr\left[\delta_{j}>\left(1+\sqrt{\frac{\log\alpha}{\alpha}}\right)\mathbb{E}\left[\delta_{j}\right]\right]\\
 & \le\exp\left(-\frac{\left(\sqrt{\frac{\log\alpha}{\alpha}}\right)^{2}}{2+\sqrt{\frac{\log\alpha}{\alpha}}}\cdot2\alpha\right)\\
 & =\exp\left(-\frac{\log\alpha}{2+\sqrt{\frac{\log\alpha}{\alpha}}}\right)\\
 & =\exp\left(-\Omega\left(\log\alpha\right)\right)\\
 & =o\left(\alpha^{-1}\right).
\end{align*}
where the second inequality follows from Lemma \ref{lem:tech kcs mul}.
Since $\alpha=k^{0.4}$, we have $\Pr\left[\delta_{j}>d\mid j\in R_{1}\right]\le o\left(1\right)$,
so the lemma holds.
\end{proof}
\begin{lem}[\cite{Brubach2019}, Lemma 1]
\label{lem:kcs coloring}Given a directed graph $G=\left(V,E\right)$
with maximum out-degree at most $d$, there exists a polynomial-time
algorithm that finds a coloring $\chi$ of $G$’s undirected version
such that $\left|\chi\right|$, the number of colors used by $\chi$,
is at most $2d+1$.
\end{lem}
\begin{proof}
The proof is unaffected by the change made to the CRS and remains
the same as in \cite{Brubach2019} (see Lemma 1).
\end{proof}
\begin{rem*}
Note that the bound on the number of colors given in Lemma \ref{lem:kcs coloring}
is tight. Indeed, for any $d\in\mathbb{Z}_{\ge1}$, consider $G=K_{2d+1}$
--- the complete graph on $2d+1$ vertices. Label the vertices with
non-negative integers modulo $2d+1$, and for every vertex $i\in\left[2d\right]\cup\left\{ 0\right\} $
orient the edges $\left(i,i+1\right)$, $\dots$, $\left(i,i+d\right)$
away from $i$ and the edges $\left(i,i+d+1\right)$, $\dots$, $\left(i,i+2d\right)$
towards $i$. The resulting directed graph has maximum out-degree
$d$, while any proper coloring of the original undirected graph uses
$2d+1$ colors. Moreover, the directed graph can be obtained in step
3 of Algorithm \ref{alg:kcs crs} for a $2d$-CS-PIP with $2d+1$
variables corresponding to vertices and $\left(2d+1\right)\cdot d$
constraints corresponding to edges.
\end{rem*}
\begin{rem*}
Note that there is a significant gap between the lower bound on the
correlation gap of the natural LP relaxation of $k$-CS-PIP given
by the $\frac{1}{4k+o\left(k\right)}$-balanced CRS we presented in
Algorithm \ref{alg:kcs crs} and the upper bound $\frac{1}{2\left(k-1+\frac{1}{k}\right)}$
following from Example \ref{exa:kcs nat} by Lemma \ref{lem:cg ub by ig}.
Judging by our analysis, manipulating the parameters $\alpha$, $\ell$,
and $d$ in Algorithm \ref{alg:kcs crs} can lead to changes in $o\left(k\right)$,
but not to an improvement in the leading term. A slight refinement
of the upper bound is possible by directly computing the correlation
gap of Example \ref{exa:kcs nat}: since any integral solution contains
at most one item, we have
\[
\sigma_{v}\left(R\right)=\begin{cases}
1, & R\neq\emptyset,\\
0, & R=\emptyset,
\end{cases}
\]
and thus 
\[
\CG\left(a,v,x\right)=\frac{1-\left(1-\frac{2-\varepsilon}{k}\right)^{n}}{\left(2-\varepsilon\right)\cdot\left(k-1+\frac{1}{k}\right)}\xrightarrow{\varepsilon\to+0,n\to\infty}\frac{1-\exp\left(-\frac{2}{k}\right)}{2\left(k-1+\frac{1}{k}\right)}.
\]
\end{rem*}

\newpage{}

\section{Conclusion and outlook}

In Section \ref{sec:hg}, we presented a monotone $\frac{1-e^{-k}}{k}$-balanced
CRS for the fractional hypergraph matching polytope with $\left|e\right|\le k$
for every hyperedge $e$. Since our algorithm is a direct generalization
of the monotone $\frac{1-e^{-2}}{2}$-balanced CRS for the fractional
matching polytope (for general graphs) from \cite{Bruggmann2019},
which is non-optimal, we believe that our scheme is non-optimal as
well. Schemes with stronger balancedness might be obtained by utilizing
more information about the various matching polytopes, e.g.,\,by
exploiting the odd set constraints of the integral matching polytope
for general graphs. This follows from the observation that the proof
of Theorem \ref{thm:hg gen} only uses the degree constraints (as
do the proofs of results in \cite{Bruggmann2019}).

One might also consider a direct generalization of the optimal monotone
CRS for the bipartite matching polytope from \cite{Bruggmann2019}
to the $k$-partite hypergraph matching setting in the same fashion
as for matchings in general graphs and hypergraphs. However, this
generalization does not go through. More specifically, if we simply
modify step 2 of Algorithm \ref{alg:hg gen crs} to assign 
\[
\left(y_{x}^{R}\right)_{e}\coloneqq\frac{q_{e}}{\max\left\{ \sum_{f\in\delta\left(v\right)}q_{f}\mid v\in e\right\} },
\]
then we lose the correctness of our contention resolution scheme.
This can be seen from the following example: take the graph $K_{3}$
as a base and add an isolated vertex ``in the middle'' of every
edge, thus forming a $3$-partite hypergraph with $6$ vertices and
$3$ hyperedges, each consisting of $3$ vertices and sharing one
distinct vertex with each of the other two hyperedges. For $x=\frac{1}{3}\cdot1_{3}$
and $R=\left[3\right]$ as the input, with non-zero probability the
algorithm sets $q=1_{3}$ and outputs $y_{x}^{R}=\frac{1}{2}\cdot1_{3}$,
which lies outside the integral matching polytope of our hypergraph.
Since in \cite{Bruggmann2019} the monotone CRS for the bipartite
matching polytope is used as a subroutine in the improved monotone
CRS for the general matching polytope, a direct generalization of
the latter is also not possible.

In Section \ref{subsec:klp ck}, we proved that the correlation gap
of the fractional knapsack polytope for class-$k$ instances is at
least $\frac{1-e^{-2}}{2}$. There is strong evidence indicating that
our analysis can be improved for any $k\ge2$ (see Conjecture \ref{conj:klp simp})
and that the class-$k$ instances presented in Example \ref{exa:klp ck tight cg}
attain the lower bound on the correlation gap for their respective
class. Based on Example \ref{exa:klp ck tight cg} and the contention
resolution schemes presented in Section \ref{subsec:klp big small},
we conjecture that the correlation gap for the fractional knapsack
polytope in the general case (without restrictions on item sizes)
is $\frac{1-e^{-2}}{2}$. Moreover, it seems that items of big size
and with small value of $x$ are the ``bottleneck'', i.e.,\,we
believe that the correlation gap and balancedness of contention resolution
schemes are defined by these items. Overall, we developed a good understanding
of how to handle separate classes of items. However, we still do not
how to combine them efficiently (in terms of balancedness of contention
resolution schemes): the contention resolution schemes for instances
with arbitrary item sizes presented in Section \ref{subsec:klp gen}
have balancedness far below $\frac{1-e^{-2}}{2}$. 

In Section \ref{sec:kcs}, we slightly modify an existing approximation
algorithm algorithm for $k$-CS-PIP based on the strengthened LP relaxation
from \cite{Brubach2019} to obtain a $\frac{1}{4k+o\left(k\right)}$-balanced
contention resolution scheme for the natural LP relaxation of $k$-CS-PIP.
By combining this scheme with a sampling step (where each item $j\in\left[n\right]$
is independently sampled with probability $x_{j}$), we immediately
get a $\left(4k+o\left(k\right)\right)$-approximation algorithm based
on the natural LP relaxation, which asymptotically improves the $8k$-approximation
algorithm given in \cite{Bansal2012}. However, the question about
the integrality gap of the natural LP relaxation of $k$-CS-PIP remains
open, since the best currently known lower bound is $2k-1$, attained
by an example from \cite{Bansal2012} (see Example \ref{exa:kcs nat}
in Appendix \ref{subsec:ig kcs}). Judging by our analysis, it seems
that the balancedness of the CRS cannot be improved beyond the $o\left(k\right)$
term without significantly modifying the scheme. We believe that this
scheme (as well as the one in \cite{Brubach2019}) is not monotone
due to the complex interaction between removal of items causing tiny,
medium, and big blocking events and different stages of the algorithm.
Potentially, a contention resolution scheme for the knapsack polytope
could be extended to the $k$-column sparse setting, although the
standard technique mentioned in Section \ref{subsec:intro cg} yields
a scheme whose balancedness is too weak.

\newpage{}

\bibliographystyle{plain}
\phantomsection\addcontentsline{toc}{section}{\refname}\bibliography{bibliography}

\newpage{}

\appendix

\section{\label{sec:tech hg}Technical statements used in Section \ref{sec:hg}}

In Section \ref{sec:hg}, we use the following well-known properties
of random variables.
\begin{lem}
\label{lem:tech exp rvs}Let $\xi\in\mathbb{R}_{>0}$ and $\eta\in\mathbb{R}_{>0}$
be positive parameters, and let $\Exp_{1}\left(\xi\right)$ and $\Exp_{2}\left(\eta\right)$
be two independent exponential random variables. Then the following
two properties hold:
\begin{enumerate}
\item $\min\left\{ \Exp_{1}\left(\xi\right),\Exp_{2}\left(\eta\right)\right\} \sim\Exp\left(\xi+\eta\right)$;
\item $\Pr\left[\Exp_{1}\left(\xi\right)<\Exp_{2}\left(\eta\right)\right]=\frac{\xi}{\xi+\eta}$.
\end{enumerate}
\end{lem}
\begin{lem}
\label{lem:tech pois rvs}Let $\xi\in\mathbb{R}_{>0}$ and $\eta\in\mathbb{R}_{>0}$
be positive parameters, let $\Pois_{1}\left(\xi\right)$ and $\Pois_{2}\left(\eta\right)$
be two independent Poisson random variables. For any $\ell\in\mathbb{Z}_{\ge1}$,
let $\Bern_{i}\left(\frac{\xi}{\ell}\right)$, $i\in\left[\ell\right]$,
be a set of independent Bernoulli random variables. Then the following
two properties hold:
\begin{enumerate}
\item $\Pois_{1}\left(\xi\right)+\Pois_{2}\left(\eta\right)\sim\Pois\left(\xi+\eta\right)$;
\item $\lim_{\ell\to\infty}\sum_{i=1}^{\ell}\Bern_{i}\left(\frac{\xi}{\ell}\right)\sim\Pois\left(\xi\right)$.
\end{enumerate}
\end{lem}

\section{\label{sec:tech klp}Technical statements used in Section \ref{sec:klp}}

Here we provide technical statements used in Section \ref{sec:klp}.
\begin{lem}[Analog of Theorem 2 from \cite{Carr2002}]
\label{lem:tech klp dom}Given an LP relaxation $P\subseteq\mathbb{R}_{\ge0}^{n}$
of an integer polytope $Z\subseteq\mathbb{R}_{\ge0}^{n}$, an $r$-approximation
heuristic algorithm $\mathcal{A}$ to a max-IP problem with $P$ as
the feasible region, and any solution $x^{*}\in P$, there exists
a polytime algorithm that finds a polynomial number of non-negative
coefficients $\lambda\in\mathbb{R}_{\ge0}^{k}$ and integral solutions
$z_{1},\dots,z_{k}\in Z$ such that $\sum_{i\in\left[k\right]}\lambda_{i}=1$
and 
\[
\frac{x^{*}}{r}\le\sum_{i\in\left[k\right]}\lambda_{i}z_{i}.
\]
\end{lem}
Recall that in Section \ref{sec:klp} we denote the cumulative distribution
function of a $\Binom\left(n,p\right)$ random variable as
\[
B\left(s;n,p\right)\coloneqq\sum_{j=0}^{s}\binom{n}{j}p^{j}\left(1-p\right)^{n-j}\quad\forall n\in\mathbb{Z}_{\ge1},\ p\in\left[0,1\right],\ s\in\left[n\right]\cup\left\{ 0\right\} 
\]
and the cumulative distribution function of a $\Pois\left(\lambda\right)$
random variable as
\[
P\left(s;\lambda\right)\coloneqq\sum_{j=0}^{s}\frac{\lambda^{j}e^{-\lambda}}{j!}\quad\forall\lambda\in\mathbb{R}_{>0},\ s\in\mathbb{Z}_{\ge0}.
\]
In addition, we define
\[
F\left(s,\lambda\right)\coloneqq\frac{1}{\lambda}\left(s-\sum_{j=0}^{s-1}P\left(j;\lambda\right)\right)\quad\forall s\in\mathbb{Z}_{\ge1},\ \lambda\in\left[0,s+1\right].
\]

\begin{figure}[h]
\caption{\label{fig:tech klp rlbf}Numerical evidence for inequalities (\ref{eq:klp simp ff 1})
and (\ref{eq:klp simp ff 2}) in Conjecture \ref{conj:klp simp} (see
observation after Lemma \ref{lem:tech klp f lb}).}

\noindent \centering{}\includegraphics[width=0.6\columnwidth,type=eps,natwidth=478,natheight=206]{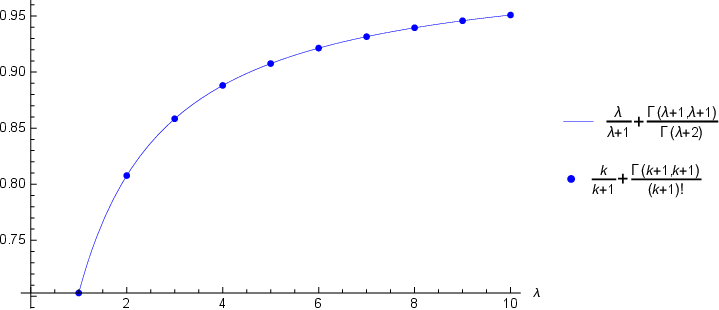}
\end{figure}

\begin{lem}[\cite{Anderson1967}, Corollary 2.1]
\label{lem:tech klp cdf}If $n\ge\lambda$ and $s\le\lambda-1$,
then $B\left(s;n,\frac{\lambda}{n}\right)\le P\left(s;\lambda\right)$.
\end{lem}
\begin{lem}
\label{lem:tech klp f dec}For any $k\in\mathbb{Z}_{\ge1}$ and any
$0\le\lambda_{1}\le\lambda_{2}\le k+1$, we have $F\left(k,\lambda_{1}\right)\ge F\left(k,\lambda_{2}\right)$.
\end{lem}
\begin{proof}
Let $k\in\mathbb{Z}_{\ge1}$ and $\lambda\in\left[0,k+1\right]$.
Then
\begin{align*}
F\left(k,\lambda\right) & =\frac{1}{\lambda}\left(k-\sum_{j=0}^{k-1}P\left(j;\lambda\right)\right)\\
 & =\frac{1}{\lambda}\left(k-\sum_{j=0}^{k-1}\left(k-j\right)\frac{\lambda^{j}e^{-\lambda}}{j!}\right)\\
 & =\frac{1}{\lambda}\left(k-k\sum_{j=0}^{k-1}\frac{\lambda^{j}e^{-\lambda}}{j!}+\sum_{j=1}^{k-1}\frac{\lambda^{j}e^{-\lambda}}{\left(j-1\right)!}\right)\\
 & =\frac{1}{\lambda}\left(k-k\sum_{j=0}^{k-1}\frac{\lambda^{j}e^{-\lambda}}{j!}+\lambda\sum_{j=0}^{k-2}\frac{\lambda^{j}e^{-\lambda}}{j!}\right)\\
 & =\frac{1}{\lambda}\left(k+\left(\lambda-k\right)\sum_{j=0}^{k-1}\frac{\lambda^{j}e^{-\lambda}}{j!}-\frac{\lambda^{k}e^{-\lambda}}{\left(k-1\right)!}\right)\\
 & =\frac{1}{\lambda}\left(k+\left(\lambda-k\right)\frac{\Gamma\left(k,\lambda\right)}{\left(k-1\right)!}-\frac{\lambda^{k}e^{-\lambda}}{\left(k-1\right)!}\right),
\end{align*}
where $\Gamma\left(k,\lambda\right)$ is the incomplete gamma function,
and the last equality holds since
\[
\Gamma\left(k,\lambda\right)=\left(k-1\right)!\cdot\sum_{s=0}^{k-1}\frac{\lambda^{s}e^{-\lambda}}{s!}\quad\forall k\in\mathbb{Z}_{\ge1}.
\]
Taking the partial derivative by $\lambda$, we get:
\begin{align*}
\frac{\partial F\left(k,\lambda\right)}{\partial\lambda} & =-\frac{1}{\lambda^{2}}\left(k+\left(\lambda-k\right)\frac{\Gamma\left(k,\lambda\right)}{\left(k-1\right)!}-\frac{\lambda^{k}e^{-\lambda}}{\left(k-1\right)!}\right)\\
 & +\frac{1}{\lambda}\left(\frac{\Gamma\left(k,\lambda\right)}{\left(k-1\right)!}-\left(\lambda-k\right)\frac{\lambda^{k-1}e^{-\lambda}}{\left(k-1\right)!}-\left(k-\lambda\right)\frac{\lambda^{k-1}e^{-\lambda}}{\left(k-1\right)!}\right)\\
 & =-\frac{1}{\lambda^{2}}\left(k-k\frac{\Gamma\left(k,\lambda\right)}{\left(k-1\right)!}-\frac{\lambda^{k}e^{-\lambda}}{\left(k-1\right)!}\right)\\
 & =-\frac{1}{\lambda^{2}}\left(k-\frac{\Gamma\left(k+1,\lambda\right)}{\left(k-1\right)!}\right)\le0,
\end{align*}
where the first equality holds, since 
\[
\frac{\partial\Gamma\left(k,\lambda\right)}{\partial\lambda}=-\lambda^{k-1}e^{-\lambda},
\]
the third equality holds, since 
\[
k\Gamma\left(k,\lambda\right)+\lambda^{k}e^{-\lambda}=\Gamma\left(k+1,\lambda\right),
\]
and the last inequality holds, since
\[
\Gamma\left(k+1,\lambda\right)=\int_{\lambda}^{\infty}y^{k}e^{-y}dy\le\int_{0}^{\infty}y^{k}e^{-y}dy=\Gamma\left(k+1\right)=k!.
\]
Thus, $F\left(k,\lambda\right)$ is a monotonically decreasing function
of $\lambda\in\left[0,k+1\right]$ for any fixed $k\in\mathbb{Z}_{\ge1}$,
so the lemma holds.
\end{proof}
\begin{lem}
\label{lem:tech klp f lb}For any $k\in\mathbb{Z}_{\ge1}$, we have
$F\left(k,k+1\right)\ge\frac{1-e^{-2}}{2}$.
\end{lem}
\begin{proof}
Note that for any $\lambda\in\left[0,k+1\right]$ we have
\begin{align*}
F\left(k,\lambda\right) & =\frac{1}{\lambda}\left(k+\left(\lambda-k\right)\frac{\Gamma\left(k,\lambda\right)}{\left(k-1\right)!}-\frac{\lambda^{k}e^{-\lambda}}{\left(k-1\right)!}\right)\\
 & =\frac{1}{\lambda}\left(k+\left(\lambda-k\right)\frac{k\Gamma\left(k,\lambda\right)}{k!}-\frac{\lambda^{k}e^{-\lambda}}{\left(k-1\right)!}\right)\\
 & =\frac{1}{\lambda}\left(k+\left(\lambda-k\right)\frac{\Gamma\left(k+1,\lambda\right)}{k!}-\left(\lambda-k\right)\frac{\lambda^{k}e^{-\lambda}}{k!}-\frac{\lambda^{k}e^{-\lambda}}{\left(k-1\right)!}\right)\\
 & =\frac{1}{\lambda}\left(k+\left(\lambda-k\right)\frac{\Gamma\left(k+1,\lambda\right)}{k!}-\frac{\lambda^{k+1}e^{-\lambda}}{k!}\right),
\end{align*}
so
\[
F\left(k,k+1\right)=\frac{k}{k+1}+\frac{\Gamma\left(k+1,k+1\right)}{\left(k+1\right)!}-\frac{\left(k+1\right)^{k+1}e^{-\left(k+1\right)}}{\left(k+1\right)!}.
\]
Next, note that $\frac{\Gamma\left(k+1,k+1\right)}{\left(k+1\right)!}\ge0$
and that the function $k\mapsto\frac{k}{k+1}$ is monotonically increasing
for $k\in\mathbb{R}_{\ge0}$. Moreover, for any $k\ge2$ we have
\[
\frac{\left(k+1\right)^{k+1}e^{-\left(k+1\right)}}{\left(k+1\right)!}\le\frac{k^{k}e^{-k}}{k!},
\]
which holds since for any $k\ge0$ we have
\[
e\ge\frac{\left(k+1\right)^{k}}{k^{k}}=\left(1+\frac{1}{k}\right)^{k},
\]
Thus, for any $k\ge2$, we have
\begin{align*}
F\left(k,k+1\right) & \ge\left.\left(\frac{k}{k+1}-\frac{\left(k+1\right)^{k+1}e^{-\left(k+1\right)}}{\left(k+1\right)!}\right)\right|_{k=2}\\
 & =\frac{2}{3}-\frac{9e^{-3}}{2}>\frac{1-e^{-2}}{2},
\end{align*}
whereas for $k=1$ we have $F\left(1,2\right)=\frac{1-e^{-2}}{2}$.
\end{proof}
Note that to prove inequalities (\ref{eq:klp simp ff 1}) and (\ref{eq:klp simp ff 2})
in Conjecture \ref{conj:klp simp}, it is sufficient to show that
$\frac{k}{k+1}+\frac{\Gamma\left(k+1,k+1\right)}{\left(k+1\right)!}$
and $\frac{\lambda}{\lambda+1}+\frac{\Gamma\left(\lambda+1,\lambda+1\right)}{\Gamma\left(\lambda+2\right)}$
are monotonically increasing functions for $k\in\mathbb{Z}_{\ge1}$
and $\lambda\in\mathbb{R}_{\ge1}$, respectively. As shown in Figure
\ref{fig:tech klp rlbf}, this fact seems to hold numerically.

\section{\label{sec:tech kcs}Technical statements used in Section \ref{sec:kcs}}

In Section \ref{sec:kcs}, we use the following Chernoff--Hoeffding-type
bounds.
\begin{lem}[Multiplicative form of the Chernoff--Hoeffding bounds]
\label{lem:tech kcs mul}Let $X_{1},\dots,X_{n}\in\left[0,1\right]$
be independent random variables such that $\mathbb{E}\left[\sum_{i\in\left[n\right]}X_{i}\right]\le\mu$.
Then for any $\delta>0$, we have
\[
\Pr\left[\sum_{i\in\left[n\right]}X_{i}\ge\left(1+\delta\right)\mu\right]\le\left(\frac{e^{\delta}}{\left(1+\delta\right)^{\left(1+\delta\right)}}\right)^{\mu}\le\exp\left(-\frac{\delta^{2}\mu}{2+\delta}\right).
\]
\end{lem}
\begin{lem}[Chefnoff--Hoeffding bounds]
\label{lem:tech kcs pow}Let $X_{1},\dots,X_{n}\in\left[0,1\right]$
be independent random variables and suppose that $\mathbb{E}\left[\sum_{i\in\left[n\right]}X_{i}\right]\le c_{1}\le c_{2}$
for some $c_{1},c_{2}\in\mathbb{R}_{>0}$ . Then
\[
\Pr\left[\sum_{i\in\left[n\right]}X_{i}\ge c_{2}\right]\le\left(\frac{c_{1}e}{c_{2}}\right)^{c_{2}}.
\]
\end{lem}

\section{\label{sec:ig exas}Bounds and examples for integrality gaps}

In this section, we give a summary of examples and bounds for integrality
gaps of LP relaxations for various problems.

\subsection{\label{subsec:ig hg}Hypergraph matching}

Now consider the hypergraph matching problem for hypergraphs with
$\left|e\right|\le k$ for every hyperedge $e$. Füredi \cite{Fueredi1981}
shows that a finite projective plane of order $n=k-1$ (if it exists)
gives an example with integrality gap $k-1+\frac{1}{k}$, which is
tight in the unweighted case (as proved in \cite{Fueredi1993}). Indeed,
a finite projective plane of order $n$ consists of $n^{2}+n+1$ points
and $n^{2}+n+1$ lines such that each line goes through $n+1$ points
and there are $n+1$ lines going through each point. Thus, treating
the finite projective plane as a hypergraph, we see that any matching
consists of at most $1$ hyperedge, while $x=\frac{1}{n+1}\cdot1_{n^{2}+n+1}$
is a feasible fractional solution of value $\left(n^{2}+n+1\right)\cdot\left(\frac{1}{n+1}\right)=n+\frac{1}{n+1}$.
Therefore, for certain values of $k$, a finite projective plane of
order $n=k-1$ (if it exists) gives an example with integrality gap
$k-1+\frac{1}{k}$.

Füredi \cite{Fueredi1981} ties the difficulty of determining the
exact integrality gap of the hypergraph matching problem to the fact
that it is problematic to decide if there exists a finite projective
plane of an arbitrary order $n$. It is well-known that if $n$ is
a power of a prime, then there is a projective plane of order $n$.
By the Bruck--Ryser--Chowla theorem \cite{Bruck1949,Chowla1950},
if $n\equiv1\left(\mod4\right)$ or $n\equiv2\left(\mod4\right)$,
then a projective plane of order $n$ exists only if $n$ is a sum
of two squares, which rules out the case $n=6$. The case $n=10$
is ruled out by a massive computer search \cite{Lam1991}, and the
case $n=12$ remains open, as well as the general case. Thus, constructing
an example attaining the integrality gap $k-1+\frac{1}{k}$ (or with
an integrality gap arbitrarily close to $k-1+\frac{1}{k}$) for an
arbitrary value of $k$ remains an open problem.

\subsection{\label{subsec:ig klp}Knapsack problem}

It is well-known that the integrality gap of the natural LP relaxation
of the knapsack problem is $2$ an that the greedy algorithm gives
a $2$-approximation. In this section, we consider instances where
every item $j\in\left[n\right]$ has size $a_{j}\le\frac{1}{2}$ and
class-$k$ instances, which we introduced in Section \ref{subsec:klp ck},
and prove a stronger integrality gap of the natural LP relaxation
and a stronger approximation ratio for the greedy algorithm in both
of these cases.
\begin{lem}
\label{lem:ig klp small}The integrality gap of any instance of the
knapsack problem where every item $j\in\left[n\right]$ has size $a_{j}\le\frac{1}{2}$
is at most $\frac{3}{2}$.
\end{lem}
\begin{proof}
Let $n\in\mathbb{Z}_{\ge1}$, $a\in\left[0,\frac{1}{2}\right]^{n}$,
and $v\in\mathbb{R}_{>0}^{n}$. Without loss of generality, we may
assume that $\frac{v_{i_{1}}}{a_{i_{1}}}\ge\frac{v_{i_{2}}}{a_{i_{2}}}$
for any $1\le i_{1}\le i_{2}\le n$. Let $x\in P\left(a\right)$ be
the optimal fractional solution of our class-$k$ knapsack instance.
As is well-known, $x$ can be obtained using the greedy algorithm,
so for some $m\in\left[n\right]$ we have
\[
x_{i}=\begin{cases}
1, & i\in\left[m-1\right],\\
\frac{1-\sum_{i\in\left[m-1\right]}a_{i}}{a_{m}}, & i=m,\\
0, & i\in\left\{ m+1,\dots,n\right\} .
\end{cases}
\]
To simplify notation, denote $w\coloneqq\sum_{i\in\left[n\right]}v_{i}x_{i}$.
Our goal is to show that there exists an integral solution supported
on $\left[m\right]$ of value at least $\frac{2}{3}w$.

Without loss of generality, we may assume that $x_{m}\in\left(0,1\right)$,
since otherwise $x$ is an integral solution of value $w\ge\frac{2}{3}w$.
Note that if $\sum_{i\in\left[m-1\right]}v_{i}\ge\frac{2}{3}w$ or
$v_{m}\ge\frac{2}{3}w$, then $\chi^{\left[m-1\right]}$ or $\chi^{\left\{ m\right\} }$
is an integral solution of value at least $\frac{2}{3}w$. Thus, we
may assume that $\sum_{i\in\left[m-1\right]}v_{i}<\frac{2}{3}w$ and
$v_{m}<\frac{2}{3}w$, which can be rewritten together as $\frac{2}{3}w>v_{m}\ge v_{m}x_{m}>\frac{1}{3}w$.
Since 
\[
w=\sum_{i\in\left[n\right]}\frac{v_{i}}{a_{i}}\cdot a_{i}x_{i}\ge\frac{v_{m}}{a_{m}}\cdot\sum_{i\in\left[n\right]}a_{i}x_{i}=\frac{v_{m}}{a_{m}},
\]
we have $a_{m}x_{m}>\frac{1}{3}$, so $a_{m}>\frac{1}{3}$ (as $x_{m}\le1$)
and $x_{m}>\frac{2}{3}$ (as $a_{m}\le\frac{1}{2}$). Thus, the extra
space needed to add the item $m$ to the knapsack is $\left(1-x_{m}\right)a_{m}<\frac{1}{6}$.
Now, consider the following three cases.
\begin{enumerate}
\item Suppose there exist two items $j,j'\in\left[m-1\right]$ with size
at least $\frac{1}{6}$. Then remove the item with smaller value,
say $j'$, and add $m$. Note that $v_{j'}<\frac{1}{3}w$, since otherwise
\[
\sum_{i\in\left[m-1\right]}v_{i}\ge v_{j}+v_{j'}\ge\frac{2}{3}w,
\]
which was ruled out above. Thus, $\chi^{\left[m\right]\setminus\left\{ j'\right\} }$
is an integral solution of value at least $\sum_{i\in\left[m\right]}v_{i}-\frac{1}{3}w>w-\frac{1}{3}w=\frac{2}{3}w$,
as desired.
\item Suppose there exists exactly one item $j\in\left[m-1\right]$ with
size at least $\frac{1}{6}$. If $v_{j}<\frac{1}{3}w$, then we can
replace $j$ with $m$ and get an integral solution of value at least
$\frac{2}{3}w$, as was shown in the previous case. Hence, we may
assume that $v_{j}\ge\frac{1}{3}w$, which means that $\sum_{i\in\left[m-1\right]\setminus\left\{ j\right\} }v_{i}<\frac{1}{3}w$.
Thus, $\chi^{\left\{ j,m\right\} }$ is an integral solution with
value at least $\frac{2}{3}w$, as desired.
\item Suppose that all items in $\left[m-1\right]$ have size strictly smaller
than $\frac{1}{6}$. Then let $R\subseteq\left[m-1\right]$ be the
minimum cardinality set that minimizes $\sum_{i\in R}v_{i}$ and such
that $\sum_{i\in R}a_{i}\ge\frac{1}{6}$. Note that $\sum_{i\in R}a_{i}<\frac{1}{3}$,
since otherwise removing any item from $R$ would reduce its cardinality
without violating the required properties. Since the item $m$ does
not fully fit into the knapsack, we have $\sum_{i\in\left[m-1\right]}a_{i}>\frac{1}{2}$.
Thus, we have $\sum_{i\in\left[m-1\right]\setminus R}a_{i}>\frac{1}{2}-\sum_{i\in R}a_{i}>\frac{1}{6}$,
which implies that $\sum_{i\in\left[m-1\right]\setminus R}v_{i}\ge\sum_{i\in R}v_{i}$
by construction of $R$. This means that $\sum_{i\in R}v_{i}<\frac{1}{3}w$,
since otherwise we have $\sum_{i\in\left[m-1\right]}v_{i}\ge\frac{2}{3}w$.
Thus, $\chi^{\left[m\right]\setminus\left\{ R\right\} }$ is an integral
solution of value at least $\frac{2}{3}w$, as desired.
\end{enumerate}
\end{proof}
\begin{lem}
\label{lem:ig klp ck}The integrality gap of any class-$k$ instance
of the knapsack problem is at most $\frac{k+1}{k}$.
\end{lem}
\begin{proof}
Let $k\in\mathbb{Z}_{\ge1}$, $n\in\mathbb{Z}_{\ge1}$, $a\in\left(\frac{1}{k+1},\frac{1}{k}\right]^{n}$,
and $v\in\mathbb{R}_{>0}^{n}$. If $n\le k$, then $P\left(a\right)=P_{I}\left(a\right)$,
and the integrality gap is $1$. Thus, going forward we assume that
$n>k$. In addition, without loss of generality, we may assume that
$\frac{v_{i_{1}}}{a_{i_{1}}}\ge\frac{v_{i_{2}}}{a_{i_{2}}}$ for any
$1\le i_{1}\le i_{2}\le n$. Let $x\in P\left(a\right)$ be the optimal
fractional solution of our class-$k$ knapsack instance. It is well-known
that $x$ can be obtained using the greedy algorithm, so we can write
$x$ explicitly:
\[
x_{i}=\begin{cases}
1, & i\in\left[k\right],\\
\frac{1-\sum_{i\in\left[k\right]}a_{i}}{a_{k+1}}, & i=k+1,\\
0, & i\in\left\{ k+2,\dots,n\right\} .
\end{cases}
\]
Note that $x_{k+1}\in\left[0,1\right)$, since $a\in\left(\frac{1}{k+1},\frac{1}{k}\right]^{n}$.
Now, consider the integral solution $\bar{x}\in P_{I}\left(a\right)$
that contains the $k$ highest valued items among the first $k+1$.
To simplify notation, let $w\coloneqq\min\left\{ v_{i}\mid i\in\left[k+1\right]\right\} $.
Then
\[
\frac{\sum_{i\in\left[n\right]}v_{i}x_{i}}{\sum_{i\in\left[n\right]}v_{i}\bar{x}_{i}}=\frac{\sum_{i\in\left[k\right]}v_{i}+v_{k+1}x_{k+1}}{\sum_{i\in\left[k+1\right]}v_{i}-w}=1+\frac{w-v_{k+1}\left(1-x_{k+1}\right)}{\sum_{i\in\left[k+1\right]}v_{i}-w}\le1+\frac{1}{k},
\]
where the last inequality holds, since $v_{k+1}\left(1-x_{k+1}\right)\ge0$
and $\sum_{i\in\left[k+1\right]}v_{i}\ge\left(k+1\right)w$. Therefore,
the integrality gap is at most $\frac{k+1}{k}$.
\end{proof}
Note that the upper bounds proved in Lemmas \ref{lem:ig klp small}
and \ref{lem:ig klp ck} are, in fact, tight, as shown by the following
example.
\begin{example}
Consider an instance with $n\ge k+1$ items, $v=1_{n}$ and $a=\left(\frac{1}{k+1}+\varepsilon\right)\cdot1_{n}$
where $\varepsilon>0$ is arbitrarily small. Then $x=n^{-1}\left(\frac{1}{k+1}+\varepsilon\right)^{-1}\cdot1_{n}$
is a feasible fractional solution of value $\left(\frac{1}{k+1}+\varepsilon\right)^{-1}=\frac{k+1}{1+\varepsilon k}$,
while any integral solution consists of at most $k$ items and thus
has value at most $k$. Therefore, the integrality gap of this example
is 
\[
\frac{k+1}{\left(1+\varepsilon k\right)k}\xrightarrow{\varepsilon\to+0}\frac{k+1}{k}.
\]
In particular, for $k=2$, this an instance consists of items of size
$\frac{1}{3}+\varepsilon$ and has integrality gap that is arbitrarily
close to $\frac{3}{2}$.
\end{example}

\subsection{\label{subsec:ig md klp}Multi-dimensional knapsack problem}

The $d$-dimensional knapsack problem is an integer problem of the
form
\begin{align*}
\max\  & \sum_{j\in\left[n\right]}v_{j}x_{j}\\
\text{s.t.}\  & \sum_{j\in\left[n\right]}a_{ij}x_{j}\le1\quad\forall i\in\left[d\right],\\
 & x\in\left\{ 0,1\right\} ^{n},
\end{align*}
where $v\in\mathbb{R}_{\ge0}^{n}$ are item values and every column
$a_{\cdot j}\in\left[0,1\right]^{d}$ of the constraint matrix represents
the sizes of the item $j\in\left[n\right]$ in each of the $d$ knapsack
constraints. The natural LP relaxation is obtained by replacing the
integrality constraint $x\in\left\{ 0,1\right\} ^{n}$ by $x\in\left[0,1\right]^{n}$.
\begin{lem}
The integrality gap of a $d$-dimensional knapsack problem is at most
$d+1$.
\end{lem}
\begin{proof}
Let $x$ be an optimal vertex solution of the natural LP relaxation
of a $d$-dimensional knapsack problem. Then at least $n$ constraints
are tight for $x$, at most $d$ of which can be knapsack constraints.
Thus, $x$ has at least $n-d$ integral components and at most $d$
fractional components. Without loss of generality, we may assume that
$x\neq0$. To simplify notation, let $S\coloneqq\left\{ i\in\left[n\right]\mid x_{i}=1\right\} $,
$u\coloneqq\sum_{i\in S}v_{i}$, $T\coloneqq\left\{ i\in\left[n\right]\mid x_{i}\in\left(0,1\right)\right\} $,
$w\coloneqq\max\left\{ v_{i}\mid i\in T\right\} $, and let $j\in T$
be the item with $v_{j}=w$ (given that $T\neq\emptyset$). Now, consider
the following three cases.
\begin{enumerate}
\item Suppose that $T=\emptyset$ and $S\neq\emptyset$. Then for the integral
solution $\bar{x}\coloneqq\chi^{S}$ we have
\[
\frac{\sum_{i\in\left[n\right]}v_{i}x_{i}}{\sum_{i\in\left[n\right]}v_{i}\bar{x}_{i}}=\frac{u}{u}=1.
\]
\item Suppose that $S=\emptyset$ and $T\neq\emptyset$. Then for the integral
solution $\bar{x}\coloneqq\chi^{\left\{ j\right\} }$ we have
\[
\frac{\sum_{i\in\left[n\right]}v_{i}x_{i}}{\sum_{i\in\left[n\right]}v_{i}\bar{x}_{i}}=\frac{\sum_{i\in T}v_{i}x_{i}}{w}\le\frac{\left|T\right|\cdot w}{w}\le d.
\]
\item Suppose that $S\neq\emptyset$ and $T\neq\emptyset$. Then for the
integral solution
\[
\bar{x}\coloneqq\begin{cases}
\chi^{S}, & \text{if }u\ge w,\\
\chi^{\left\{ j\right\} }, & \text{if }u<w
\end{cases}
\]
we have
\[
\frac{\sum_{i\in\left[n\right]}v_{i}x_{i}}{\sum_{i\in\left[n\right]}v_{i}\bar{x}_{i}}=\frac{u+\sum_{i\in T}v_{i}x_{i}}{\max\left\{ u,w\right\} }\le\frac{\left(\left|T\right|+1\right)\cdot\max\left\{ u,w\right\} }{\max\left\{ u,w\right\} }\le d+1.
\]
\end{enumerate}
Thus, in each of the three cases the integrality gap is at most $d+1$,
which proves the lemma.
\end{proof}
\begin{rem*}
In fact, the bound $\frac{d}{d+1}$ on the integrality gap of class-$k$
knapsack instances is tight. Indeed, consider an instance with $n=d+1$
items, $v=1_{d+1}$ and 
\[
a_{ij}=\begin{cases}
1-\varepsilon, & i\in\left[d\right],\ j=i,\\
2\varepsilon, & i\in\left[d\right],\ j\in\left[d+1\right]\setminus\left\{ i\right\} 
\end{cases}
\]
where $\varepsilon>0$ is arbitrarily small. Then the optimal fractional
solution $x$ is given by 
\[
x_{i}=\begin{cases}
\frac{1-2\varepsilon}{1+\left(2d-3\right)\varepsilon}, & i\in\left[d\right],\\
1, & i=d+1
\end{cases}
\]
and its value is
\[
\sum_{i\in\left[d+1\right]}x_{i}=\frac{d+1-3\varepsilon}{1+\left(2d-3\right)\varepsilon}\xrightarrow{\varepsilon\to+0}d+1.
\]
On the other hand, since $a_{jj}+a_{ij}>1$ for every $i\in\left[d\right]$
and $j\in\left[d+1\right]$ such that $i\neq j$, any integral solution
consists of at most one item and hence its value is at most $1$.
Thus, the integrality gap of this example goes to $d+1$ as $\varepsilon\to+0$.
\end{rem*}

\subsection{\label{subsec:ig kcs}$k$-CS-PIP}

The following example shows that the natural LP relaxation of $k$-CS-PIP
has integrality gap at least $2k-1$.
\begin{example}[\cite{Bansal2012}, page 19]
\label{exa:kcs nat}Consider the finite projective plane of order
$k-1$ where $k-1$ is a prime power, identify the items with points
of the projective plane, and set $v_{j}\coloneqq1$ and $a_{ij}\coloneqq\frac{1}{2-\varepsilon}$
for every $i\in\left[m\right]$ and $j\in\left[n\right]$, where $\varepsilon>0$
is arbitrarily small. Then setting $x_{j}=\frac{2-\varepsilon}{k}$
for every $j\in\left[n\right]$ gives a fractional LP solution of
value 
\[
\left(2-\varepsilon\right)\cdot\left(k-1+\frac{1}{k}\right)\xrightarrow{\varepsilon\to+0}2k-1,
\]
while any integral solution contains at most one item and thus has
value at most $1$.
\end{example}
The integrality gap of the strengthened LP relaxation is also at least
$2k-1$, as shown by the following example.
\begin{example}[\cite{Bansal2012}, page 19]
\label{exa:kcs str}Consider $n=m=2k-1$ items and constraints, indexing
the items modulo $n$. Set $v_{j}\coloneqq1$ for every $j\in\left[n\right]$
and
\[
a_{ij}\coloneqq\begin{cases}
1, & i=j,\\
\varepsilon, & j\in\left\{ i+1,\dots,i+k-1\right\} \left(\mod n\right),\\
0, & \text{else}
\end{cases}\quad\forall i,j\in\left[n\right].
\]
where $\varepsilon>0$ is arbitrarily small. Setting $x_{j}=1-k\varepsilon$
gives a fractional solution of value 
\[
\left(1-k\varepsilon\right)\cdot n\xrightarrow{\varepsilon\to+0}2k-1,
\]
while every integral solution contains at most one item and thus has
value at most $1$.
\end{example}

\end{document}